\documentclass[12pt,floatfix,tightenlines,superscriptaddress,aps,prb,citeautoscript,nobibnotes]{revtex4}

\usepackage{graphicx,graphics}
\usepackage{amsmath,epsf,subfigure}
\usepackage{color}
\usepackage{amsfonts,amsmath,graphics}
\usepackage{bm,subfigure}
\usepackage{graphicx,epsf}
\usepackage{amssymb}
\usepackage{afterpage}
\usepackage{float}
\usepackage{tikz-cd}
\usepackage{MnSymbol}
\usepackage{amsmath}
\usepackage{comment}
\usepackage{chemgreek}

\newcommand{\obj}{molecule}
\newcommand{\conf}{state }
\newcommand{\confs}{states }
\newcommand{\kcat}{k_{\rm cat}}
\newcommand{\rcat}{\rho_{\rm cat}}

\newcommand{\tcat}{T_{C+S\to C+P}}
\newcommand{\tspo}{T_{S\to P}}
\newcommand{\E}{\mathbb{E}}
\newcommand\reac[2]{\underset{#1}{\stackrel{#2}{\rightleftharpoons}}}

\renewcommand{\.}{{\cdot}}

\newcommand{\MM}{{\rm MM}}

\newcommand{\beq}{\begin{equation}}
\newcommand{\eeq}{\end{equation}}
\newcommand{\bea}{\begin{eqnarray}}
\newcommand{\eea}{\end{eqnarray}}

\renewcommand{\a}{\alpha}
\renewcommand{\b}{\beta}
\newcommand{\g}{\gamma}

\newcommand{\e}{\epsilon}

\renewcommand{\r}{k}

\renewcommand{\P}{\mathbb{P}}
\newcommand{\bs}{\backslash}

\newcommand{\ys}{\textcolor{black}}

\begin{document}

\title{On Kinetic Constraints That Catalysis\\ Imposes on Elementary Processes}

\author{Yann Sakref}
\affiliation{Gulliver UMR CNRS 7083, ESPCI Paris, Universit\'e PSL, 75005 Paris, France}

\author{Maitane Mu\~noz-Basagoiti}
\affiliation{Gulliver UMR CNRS 7083, ESPCI Paris, Universit\'e PSL, 75005 Paris, France}
\affiliation{Institute of Science and Technology Austria, 3400 Klosterneuburg, Austria}

\author{Zorana Zeravcic}
\affiliation{Gulliver UMR CNRS 7083, ESPCI Paris, Universit\'e PSL, 75005 Paris, France}

\author{Olivier Rivoire}
\email{olivier.rivoire@espci.fr}
\affiliation{Gulliver UMR CNRS 7083, ESPCI Paris, Universit\'e PSL, 75005 Paris, France}

\begin{abstract}
\vspace{0.1 cm}
\begin{center}
    \textbf{ABSTRACT}\\
\end{center}
Catalysis, the acceleration of product formation by a substance that is left unchanged, typically results from multiple elementary processes, including diffusion of the reactants towards the catalyst, chemical steps, and release of the products. While efforts to design catalysts are often focused on accelerating the chemical reaction on the catalyst, catalysis is a global property of the catalytic cycle that involves all processes. These are controlled by both intrinsic parameters such as the composition and shape of the catalyst, and extrinsic parameters such as the concentration of the chemical species at play.  We examine here the conditions that catalysis imposes on the different steps of a reaction cycle and the respective role of intrinsic and extrinsic parameters of the system on the emergence of catalysis by using an approach based on first-passage times. We illustrate this approach for various decompositions of a catalytic cycle into elementary steps, including non-Markovian decompositions, which are useful when the presence and nature of intermediate states are a priori unknown. Our examples cover different types of reactions and clarify the constraints on elementary steps and the impact of species concentrations on catalysis.
\end{abstract}

\maketitle
\thispagestyle{myheadings}  

\section{Introduction}

Catalysts are substances that accelerate the completion of chemical reactions without being consumed in the process. They have long been studied in chemistry for their role in the industrial production of chemical products and in biology for their role in the metabolism and regulation of living processes~\cite{chorkendorff2010}. A newer context for catalysis is provided by developments in supramolecular chemistry, DNA nanotechnology and soft matter physics~\cite{Zhang2017, Zhuo2019, Baigl2022} which yield alternative chemistries where the size, shape, valence, strength and specificity of interaction of the molecules can be tuned~\cite{Feng2013, Seeman_2017,Hueckel2021,niu2019magnetic}. In these chemistries, catalysts can be designed with a control over many parameters, which provides experimental models to study catalysis and, more generally, new contexts to engineer properties inspired by biology.

Motivated by these perspectives, we have recently taken up the task of computationally designing catalysts in the context of a chemistry where the ``atoms'' consist of spherical colloidal particles that interact via pairwise potentials~\cite{mrz}. In this chemistry, the interaction potentials are isotropic with a common shape and interaction range, but with possibly different depths. Different atoms ($A,B,\dots$) can be designed to have interactions with specific strengths ($\e_{AA}, \e_{AB},\e_{BB},\dots$). We considered catalyzing the dissociation of a dimer made of two bound particles of type $A$  by a catalyst consisting of particles of another type $B$, i.e., catalyzing the reaction $A_2\to 2A$. This problem led us to ask several general questions: how to assess the presence of catalysis? How to score the performance of a catalyst? To which extent do these questions depend on the extrinsic conditions under which catalysis is analyzed, including the volume of the vessel and the presence of multiple substrates or multiple catalysts? Is it sufficient to analyze a system consisting of a single substrate and a single catalyst, as it is computationally most convenient to do, or could catalysis arise only when sufficiently many molecules are present? What extrinsic conditions are most favorable to catalysis?

More specific questions arise when studying the different steps that catalysis typically requires, including binding of the substrate(s) to the catalyst, reaction(s) in the presence of the catalyst, and release of the product(s)~\cite{lehn95}. Some of these steps can have a simple dependence on the design parameters. For instance, the final step of the catalyzed reaction that we studied in~\cite{mrz} is the release of a single particle $A$ attached to a single particle $B$ of the catalyst which depends on the interaction strength between $A$ and $B$ ($\e_{AB}$) but not on other features of the catalyst. Since this final release must occur faster than the spontaneous reaction for catalysis to take place, the interactions between $A$ and $B$ must be weaker than the interaction between two $A$, $\e_{AB}<\e_{AA}$. It would be desirable to extend this reasoning to obtain bounds on the design parameters, but this opens up several additional questions: is it always necessary for a forward step along a catalytic cycle to be faster than the spontaneous reaction? What if we do not have the knowledge of each intermediate state, as is typically the case when starting to investigate different designs? 

Several of these questions have been previously addressed~\cite{Bowden1979,wolfenden2001depth,kozuch2012turning, vojvodic2015new,Leier2017}. Previous approaches, however, do not provide straightforward and consistent answers to all of the above questions. For instance, a well-accepted quantification of catalytic activity is the turnover number, defined as the number of substrates that a catalyst converts per unit of time~\cite{kozuch2012turning}. \ys{One limitation of the turnover number is that it does not refer to the spontaneous reaction. As a consequence, it cannot reveal if catalysis, understood as an acceleration relative to the spontaneous reaction, is indeed taking place. This is not an issue when studying substances such as enzymes, which are unambiguously accelerating reactions, but can be an issue when designing catalysts in new contexts~\cite{mrz}. The first designs are indeed likely to have limited activities, in which case establishing that catalysis is taking place is critical before considering any improvement.}

\ys{Beyond the turnover number,} enzymes are typically characterized by two distinct quantities, the catalytic rate constant $\kcat$ and the Michaelis constant $K_M$~\cite{Bowden1979}. Taken together, they account for the dependence of the reaction rate on substrate concentration $[S]$ by describing the rate of product formation per enzyme as $\kcat[S]/(K_M+[S])$, a relation known as Michaelis-Menten kinetics~\cite{Bowden1979}. By accounting for the concentration of substrates, this relationship partly accounts for extrinsic parameters. In this case, a measure of catalytic efficiency relative to the spontaneous reaction has been proposed~\cite{wolfenden2001depth}, but Michaelis-Menten kinetics rely on several assumptions that prevent its general applications~\cite{srinivasan2022guide}: the substrate must be in excess relative to the catalyst and catalysis must involve a series of states with rates of transition justifying that an intermediate complex is at a quasi-steady state. These assumptions cannot be made in general.

These limitations motivated us to develop our own criterion for defining and quantifying catalysis, a criterion from which necessary conditions for elementary steps can be derived. Our approach goes beyond assumptions that may be justified in particular cases but are not guaranteed when considering more general catalysts: (i) the spontaneous reaction must be accounted for, as catalysis refers to its acceleration~\cite{mcnaught1997compendium}; (ii) the substrate does not necessarily have to be in excess to the catalyst, as it is for instance of interest when studying some biochemical reactions~\cite{Stadtman1984,Luby-Phelps2000, Souza2013} or autocatalysis~\cite{SemenovAC}; (iii) the kinetics must not be imposed to cover both enzymes that follow Michaelis-Menten kinetics and heterogeneous catalysts that do not~\cite{laidler1982physical}, and more generally not to restrain a priori the design of new catalysts. In what follows, we present our approach and show how it addresses the different questions that we raised through analysis of several examples.

\section{Methods}

\subsection{A formal definition of catalysis}

We assume a thermal bath at a constant temperature and a reaction vessel of fixed volume. A spontaneous reaction is a thermally induced transition between two \confs that differ in their composition, i.e., where  some \obj s $S_1,\dots,S_n$ (the reactants, collectively denoted $S$) are transformed into some other \obj s $P_1,\dots, P_m$ (the products, collectively denoted $P$).
We define the time of the spontaneous reaction to be the mean time $\tspo$ to reach a final state $P$ from an initial one $S$. For instance, in the simple case of a single reactant $S$ that converts into a product $P$ at a given rate $\r_0$, the initial \conf is $S$, the final \conf is $P$, and the spontaneous reaction time is $\tspo=1/\r_0$.

The spontaneous reaction time is compared to the average time $\tcat$ to complete the reaction in the presence of an additional \obj\ $C$ in the reaction vessel. We consider that $C$ catalyses the transformation of the substrate into product if the reaction is on average faster in its presence, that is, if $\tcat<\tspo$, and if $C$ is unchanged in the process. This excludes initial and final \confs in which $C$ is not present, or is interacting with other \obj s, including substrate or product molecules. To define and quantify catalysis, we therefore propose to use the ratio $\tspo/\tcat$ both to assess the presence of catalysis, through $\tspo/\tcat>1$, and to quantify its efficiency.
\ys{This quantitative criterion comparing two average times is consistent 
with the definition of catalysis given by the International Union of Pure and Applied Chemistry (IUPAC), according to which a catalyst is ``a substance that increases the rate of a reaction"~\cite{laidler_glossary_1996}. Beyond the average time to complete a reaction, it may be of interest to consider the complete distribution of completion times, particularly when the most probable time differs from the mean time, as illustrated in Section~\ref{Conditions for catalysis with multiple $S$ and $C$}, where we compare catalysis with either one or two substrates. When considering systems with a large number of particles, however, the mean time is expected to correspond to the most likely time.}

\subsection{Decompositions in elementary steps}

As it is essential to distinguish configurations in which substrates and products are bound to the catalyst from those where they are unbound, the simplest decomposition of the catalytic process is into a cycle where configurations are partitioned into just three states: an initial state where the substrates $S$ and the catalyst $C$ are unbound, an intermediate state where they are interacting, and a final state where the products $P$ are not bound to the catalyst. For example, a model widely used in enzymology and which serves as a basis to derive Michaelis-Menten kinetics~\cite{srinivasan2021guide} is that of a Markov model with an intermediate state $CS$ that is reached from an unbound state $C+S$ at a rate $k_1$ and left either back to $C+S$ at a rate $k_{-1}$ or towards $C+P$ at a rate $k_2$, as represented by
\beq
C+S\reac{\r_{-1}}{\r_1}CS\xrightarrow[]{\r_{2}}  C+P.
\eeq
More generally, Markov models with a larger number of intermediate states are commonly introduced in studies of catalysis~\cite{Bowden1979}.

Decomposing a catalytic process as a Markov chain rests on a major assumption of separation of time scales between the time to transition between states and the time spent in the states, with local equilibration within each state so that memory of previous states is lost. Under this assumption, the transitions are quantified by rates, corresponding to exponential distributions of dwelling times within each state. But such an assumption is not necessarily valid. First, binding and unbinding events involving diffusion are described as Markov processes only under a mean-field approximation that may not be justified. In particular, the mass action kinetics that underlies the definition of a single state $C+S$ breaks down in low-dimensions~\cite{savageau1995michaelis}. 
Even in three dimensions, rates may fail to describe transitions between intermediate states when they involve the restricted diffusion of part of a molecule. Second, the number and nature of intermediate states may be unknown~\cite{mrz}.

A more general framework is that of semi-Markov processes, also known as Markov renewal processes~\cite{karlin2014first}. The only essential assumption of semi-Markov processes is a separation of time scales to define states, with dwelling times within states that are not necessarily exponentially distributed. Each state $i$ has a distribution of dwelling times $\P(\tau_i)$ in addition to probabilities $p_{i\to j}$ to transition to other states $j\neq i$ once this dwelling time is over. We will develop our formalism in this broader context. In contrast to a decomposition into a Markov process, the decomposition of a catalytic cycle into a semi-Markov process is not unique and, as such, may be more or less informative. In particular, unknown states of a Markov process may be collected in different ways in fewer states to define a coarse-grained semi-Markov process. Below, we thus present a minimal decomposition with a single intermediate state, which we denote $C\.\.S$, that allows us to give general conditions. The more states that are known, however, the more information, in the form of constraints, may be derived. 

\section{Results}

\subsection{Single molecule catalysis}

To show how these informative constraints can be derived, we start by studying a unimolecular spontaneous reaction $S\to P$ that proceeds at a rate $\r_0$. For simplicity, we first consider a closed vessel containing a single substrate $S$ and a single catalyst $C$. Applying our approach, we derive necessary and sufficient kinetic conditions on the elementary processes of the cycles for $C$ to catalyze the spontaneous reaction. The restrictions on the nature of the spontaneous reaction and the number of molecules will be lifted in the next sections.

\subsubsection{Markovian catalytic cycles}\label{sec:Markov}

Under the above assumptions, the most basic catalytic cycle comprises only one intermediate state denoted as $CS$, which is accessible from either the unbound states $C+S$ or $C+P$ and can also transition back to those states. Graphically, the cycle is represented as
\begin{equation} \label{Simple Scheme}
\begin{tikzcd}[column sep=small]
C+S \arrow[dr, shift left, "k_1"]\arrow[rr, "k_0"] & & C+P\arrow[ll, shift left, "k_{-0}"]\arrow[dl, "k_{-2}"] \\
    & CS \arrow[ul, "k_{-1}"]\arrow[ur, shift left, "k_2"]
\end{tikzcd}
\end{equation}
with $\r_{+n}$ denoting the forward rate of an elementary reaction and $\r_{-n}$ its reverse rate. Given that we consider the mean first-passage time $T_{C+S\to C+P}$ from the initial state to the absorbing state $C+P$, the transitions with rates $k_{-0}$ and $k_{-2}$ can be ignored and an equivalent representation is 
\beq
C+P\xleftarrow{k_0}C+S\reac{\r_{-1}}{\r_1}CS\xrightarrow[]{\r_{2}}  C+P,
\eeq
where $C+P$ is repeated on both sides.
The mean first-passage time from $C+S$ to $C+P$ can then be written in terms of the elementary rates (see Supporting Information~\ref{app:Markov}) as
\beq\label{eq:two}
T_{C+S\to C+P}=\frac{k_1+k_{-1}+k_2}{k_0k_{-1}+k_0k_2+k_1k_2}.
\eeq
or, to highlight constraints,
\beq\label{eq:central}
T_{C+S\to C+P}=T_{S\to P}+\rcat (T_{\rm cat}-T_{S\to P})
\eeq
with $T_{S\to P}=1/k_0$, $T_{\rm cat}=1/k_2$ and
\beq\label{eq:rhoN1}
\rcat=\frac{1}{1+(1+k_{-1}/k_2)k_0/k_1}.
\eeq
By our criterion, catalysis takes place if and only if $T_{C+S\to C+P}<T_{S\to P}$, which, by Eq.~\eqref{eq:central}, is the case if and only if $\rho_{\rm cat}>0$ and $T_{\rm cat}<T_{S\to P}$. In terms of the elementary rates, this requires that (i)~$k_1>0$, simply meaning that it is possible for the substrate to bind the catalyst, and (ii)~$k_2>k_0$, meaning that the transformation of the substrate into product on the catalyst and its release from the catalyst are overall faster than the spontaneous reaction. In particular, the presence of catalysis is independent of $k_{-1}$ and of the magnitude of $k_1$, which only needs to be non-zero. \ys{These conditions are partly counter-intuitive. Naively, one may expect that for catalysis to take place every step forward process along the catalytic cycle must be faster than the spontaneous reaction. We find here that it is not the case, and that $k_1$ may take any value. This can be understood by noting that when a putative catalyst is present, the substrate has two pathways: it can either undergo the spontaneous reaction or proceed through the potential catalytic route. If the first pathway is taken, no delay is possibly incurred and only if the second pathway is taken must we ensure that no additional time is spent relative to the spontaneous reaction ($k_2>k_0$). If this condition is not satisfied, the catalysts turns into an inhibitor that slows down the completion of the reaction. $\rho_{\rm cat}$ represents the probability to take the catalytic pathway, and, maybe also counterintuitively, this probability may be arbitrarily small, meaning that the reaction mostly occur spontaneously, without preventing catalysis to occur, although possibly only marginally.}

\ys{However, while the presence of catalysis is independent of the values of $\r_1$ and $\r_{-1}$, the efficiency of catalysis certainly depends on these quantities.} In particular, in the limit $\r_1\to 0$, catalysis becomes negligible ($\tcat\to \tspo=1/k_0$), while it is maximal in the limit $\r_1\to \infty$ ($\tcat\to T_{\rm cat}=1/k_2$). In any case, the rates $k_{-0}$ and $k_{-2}$ that describe the rates at which state $C+P$ is left are irrelevant, since we consider only the first time at which $C+P$ is reached.

The calculation can easily be extended to cases where catalysis involves a series of $N$ intermediate Markovian states of the form
\beq
C+P\xleftarrow{k_0}C+S\reac{\r_{-1}}{\r_1}CS_1\reac{\r_{-2}}{\r_2}\dots \reac{\r_{-N}}{\r_N}CS_N\xrightarrow[]{\r_{N+1}}  C+P
\eeq
In this case, we may again write
\beq\label{eq:markov}
T_{C+S\to C+P}=T_{S\to P}+\rcat (T_{\rm cat}-T_{S\to P})
\eeq
with
\beq\label{eq:Tcat_chain}
T_{\rm cat}=\sum_{n=0}^{N-1}\sum_{i=1}^{N-n}\left(\prod_{j=1}^n\frac{k_{-(i+j)}}{k_{i+j}}\right)\frac{1}{k_{i+n+1}}
\eeq
and
\beq\label{eq:chain}
\rcat=\frac{1}{1+(1+\theta k_{-1}/k_2)k_0/k_1}
\eeq
where the variables $T_{\rm cat}$ and $\theta$ are functions of all the rates $k_{\pm i}$ with $i>1$, with the exception of the rate $k_{-(N+1)}$ that describes the rate at which the final state $C+P$ is left. In particular, with one intermediate state ($N=1$), $T_{\rm cat}=1/\r_2$ and $\theta=1$, leading back to Eq.~\eqref{eq:rhoN1}. With two intermediate states ($N=2$), $T_{\rm cat}=1/\r_2+1/\r_3+\r_{-2}/(\r_2\r_3)$ and $\theta=1+\r_{-2}/\r_3$; because of the reversibility of the transitions between states, $T_{cat}$ is longer $T_{\rm cat} > 1/\r_2+1/\r_3$; more generally, $T_{\rm cat}>\sum_{i=2}^{N+1} 1/\r_i$.

From Eq.~\eqref{eq:markov}, we obtain necessary conditions for catalysis that generalize those obtained with a single intermediate state: the first step must occur at a non zero rate ($k_1>0$) and the following steps must each be faster than the spontaneous reaction
($\r_i>\r_0$ for $i\geq 2$).
While the value of $k_{-1}$ is again irrelevant for the definition of catalysis, this is not the case for the elementary rates $k_{-i}$ for $2\leq i\leq N$: the conditions $\r_i>\r_0$ for $i\geq 2$ are necessary, although not sufficient. The necessary and sufficient condition is that $T_{\rm cat}<T_{S\to P}$ and $\rho_{\rm cat} >0$, where $T_{\rm cat}$ and $\rho_{\rm cat}$ depend on the reverse rates $k_{-i}$ for $2\leq i\leq N$.

\subsubsection{Identification of favorable conditions}

Eq.~\eqref{eq:Tcat_chain} indicates that setting to zero all reverse rates $k_{-n}$ with $n>0$ effectively reduces $T_{\rm cat}$; this reflects the fact that, for catalytic cycles with a single loop, it is always favorable to operate in conditions where the backward transitions are suppressed~\cite{mrz}. Here we illustrate how this may be achieved by considering reactions that are irreversible or by removing products as they are made. 

As an illustration of the benefit of irreversible reactions, consider for instance a catalytic process with two intermediate states, $C+P\leftarrow C+S\reac{}{}CS\reac{}{}CP\to C+P$. The condition for catalysis $T_{\rm cat}<T_{S\to P}$ can be rewritten
\beq
\frac{1}{\r_{CS\to CP}}+\frac{1}{\r_{CP\to C+P}}+\frac{\r_{CP\to CS}}{\r_{CS\to CP}\r_{CP\to C+P}}
<\frac{1}{k_{S\to P}}.
\eeq
The first two terms correspond to conditions on the elementary forward rates while the third one involves an additional constraint on the reverse reaction. This third constraint vanishes when  the reverse reaction on the catalyst is irreversible, that is when $\r_{CP\to CS}=0$, which is expected when the spontaneous reaction is itself irreversible, $\r_{P\to S}=0$.

When designing a catalyst for a reaction with a forward rate $\r_{S\to P}$, we may therefore first identify necessary conditions to catalyze an irreversible reaction with the same forward rate, which is generally easier. Catalysts for the reversible reaction must indeed necessarily satisfy those conditions as well.

As an illustration of the benefit of product removal, consider a reaction $S\to P_1+P_2$ and a catalytic scheme $C+P\leftarrow C+S\reac{}{}CS\reac{}{}CP_1+P_2\to C+P_1+P_2$. 
Here the backward rate $CP_1+P_2\to CS$ can effectively be eliminated by removing $P_2$ from the system as soon as it is formed. This also makes catalysis easier and, again, a catalyst under conditions where the products remain in the reaction vessel must be a catalyst under conditions where the products are immediately removed.

These simple considerations provide insights into the conditions that are most favorable to catalysis.

\subsubsection{Beyond Markov chains} \label{section: General Criterion}

\ys{Can we extend Eq.~\eqref{eq:central} to decompositions of catalytic processes where we lack specific information about the potential configurations that $C$ and $S$ might assume upon being close to each other, or when these configurations do not constitute Markovian states, meaning we cannot assume they are independent of prior configurations?} To do so, we introduce a \ys{transitional} state $C\.\.S$ that defines a boundary between interacting and non-interacting substrate-catalyst pairs, with a catalytic process described by
\beq\label{eq:nonM}
C+P\leftarrow \{C+S\}\reac{}{} C\.\. S\to \{CS\}\to C+P.
\eeq
Here $C\.\. S$ represents a state where $C$ and $S$ are just about to interact, while $\{CS\}$ gathers all other configurations where they interact and $\{C+S\}$ those where they do not interact. Braces are introduced to indicate that these configurations do not generally define a ``state'', in the sense of a set of configurations that are equivalent as far as future transitions are concerned, as it is the case when considering a Markov chain. The arrows indicate possible transitions and $C+P$ denotes the final configurations where the product is formed and dissociated from the catalyst. In this setting, Eq.~\eqref{eq:central} still provides a decomposition of the mean first-passage time from an initial configuration $(C+S)_0$ to any of the final configurations $C+P$ but with $T_{\rm cat}$ representing the mean time to reach $C+P$ from $C\.\.S$ when taking the route via $\{CS\}$ and with $\rho_{\rm cat}=p_0q/(1-(1-q)p)$, where $p_0$ is the probability to reach $C\.\.S$ from the initial configuration $(C+S)_0$, $q$ the probability to go towards $\{CS\}$ rather than $\{C+S\}$ once in $C\.\.S$ and $p$ the probability to come back to $C\.\.S$ given that the path towards $\{C+S\}$ was taken (Supporting Information~\ref{app:semiMarkov}). Formally, the same necessary and sufficient conditions for catalysis then apply, namely $T_{\rm cat}<T_{S\to P}$ and $\rho_{\rm cat}>0$.

This generalization provides an example of necessary conditions for catalysis given a non-Markovian decomposition of the catalytic cycle. It also addresses cases where the mechanisms of catalysis are unknown, since Eq.~\eqref{eq:nonM} makes no assumption on the presence of intermediate states. In practice, $C\.\.S$ consists of configurations where $C$ and $S$ are at given distance just above their interactions range. \ys{Remarkably, we need not distinguish configurations within $\{C+S\}$ or $\{CS\}$. In particular, we need not distinguish unbound states as a function of the distance between $C$ and $S$. Instead, all the relevant information can be encapsulated into the probability and the mean time for the system to come back to $C\.\.S$ after leaving it (Supporting Information~\ref{app:semiMarkov}).} This decomposition can then be used to limit the analysis to a volume of same dimension, where configurations in $\{C+S\}$ are effectively excluded. These configurations are indeed irrelevant to the computation of $T_{\rm cat}$. This is for instance convenient in the context of molecular dynamics simulations where the time spent in $\{C+S\}$ is otherwise wasted

It is also of interest to consider situations where additional information is available that allows for a more detailed description of catalysis than Eq.~\eqref{eq:nonM}, although still not necessarily in the form of a Markov chain. For instance, catalysis may be known to follow a sequence of states $C+P\leftarrow C+S\reac{}{} CS_1\reac{}{} \cdots \reac{}{}  CS_{N}\to C+P$ described by a semi-Markov chain where the time spent in each state $C+S$ or $CS_i$ is not necessarily described by an exponential distribution, as it is the case for Markov chains. In this case, we may generalize the necessary (but not sufficient) conditions obtained for Markov chains ($k_i>k_0$ for $i\geq 2$) to $T_{CS_{i-1}\to CS_{i}\bs CS_{i-2}}<T_{S\to P}$ for each $i\geq 2$, where $CS_{i-1}\to CS_{i}\bs CS_{i-2}$ indicates that we are considering the first passage from $CS_{i-1}$ to $CS_{i}$ when accessing $CS_{i-2}$ is excluded (with $CS_0=C+S$ when $i=2$).

The previous extensions of Eq.~\eqref{eq:central} apply to a single substrate $S$, a single (candidate) catalyst $C$, and assume a spontaneous reaction $S\to P$ described by a single rate $\r_0$. Beyond this case, it does not apply without further assumptions. For instance, if the spontaneous reaction proceeds through intermediate states, no state $C\.\.S$ can generally be defined (Supporting Information~\ref{app:semiMarkov}).

Difficulties also arise when considering multi-molecular reactions, e.g., $S+S\to P$, as extending the definition of $C\.\.S$ to these reactions is not straightforward. In practice, however, it may be possible to make additional assumptions. For example, $C+S+S \reac{}{} C\. S + S \reac{}{} CS_2$ may be described by a Markov chain and only $CS_2\to \{CS_2\}\to C+P$ may require a non-Markovian description.

\subsubsection{Application to reactions with bimolecular substrates and products} \label{section: bc}

\begin{figure}[t]
\begin{center}
\includegraphics[width=.99\linewidth]{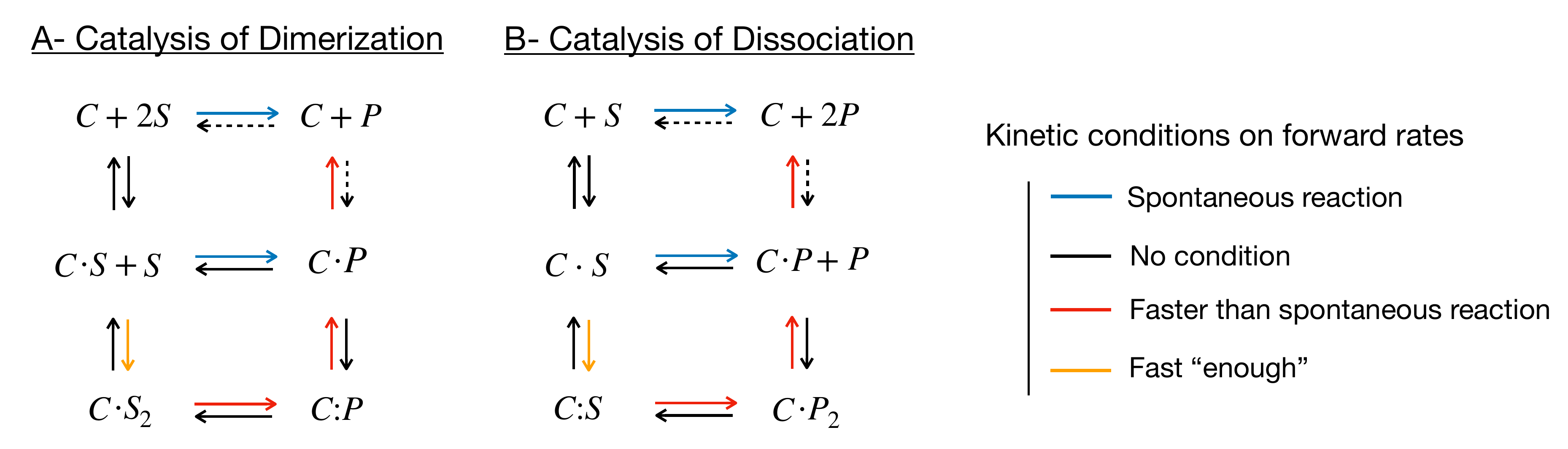}
\caption{Examples of decomposition of catalytic processes into graphs with multiple loops. {\bf A.}~Catalysis of dimer formation. {\bf B.}~Catalysis of dimer dissociation. The steps corresponding to a spontaneous reaction are indicated in blue and each forward step that must be faster than the spontaneous reaction is indicated in red. Steps indicated in orange must also be fast enough, although not necessarily faster than the spontaneous reaction.}
\label{fig:dimer}
\end{center} 
\end{figure}

Conditions on elementary steps are more subtle when catalysis is not described by a cycle with a single loop but by a graph comprising several loops. This occurs for instance when considering the spontaneous reaction the association of two monomers into a dimer or the dissociation of a dimer into two free monomers. In these cases, six states may typically be defined by the number of bonds formed between a catalyst $C$, the two monomers, and the dimer. As shown in  Fig.~\ref{fig:dimer}, these states are part of a graph that contains two loops. In the first loop, a spontaneous reaction can occur while only one particle of the substrate is bound to the catalyst, whereas the second loop describes the catalyzed reaction. 

Consider for illustration the case of the dissociation reaction (Fig.~\ref{fig:dimer}B) which we studied previously in the context of colloidal particles~\cite{mrz}. Assuming that the mean time for a spontaneous reaction is exponentially distributed (or, even more simply, that the entire process is Markovian), we can apply Eq.~\eqref{eq:central} on both loops to compute the mean time to form the product in the presence of the catalyst, $T_{C+S \rightarrow C+2P}$ (Supporting Information~\ref{app: dimer dissociation}). The necessary and sufficient conditions for catalysis take their simplest simple form if the product -the monomers in that case of a dissociation reaction- is removed upon forming. Let indeed $\rho_0$ be the probability to reach $C+2P$ from $C+S$ through $C\.S$ rather than through the direct transition $C+S\to C+2P$ and let $\rho_1$ be the probability to reach $C\.P$ from $C\.S$ through $C\:S$ rather than through the direct transition $C\.S\to C\.P+P$. We may again write $T_{C+S\to 2P}=T_{S\to 2P}+\rcat (T_{\rm cat}-T_{S\to 2P})$ with this time $\rcat=\rho_0\rho_1$ and $T_{\rm cat}=T_{C\:S\to C\.P+P\bs C\.S}+\rho_1^{-1}T_{C\.P\to C+P\bs C\.S}$. Necessary and sufficient conditions are therefore $\rho_0>0$, $\rho_1>0$ and $T_{\rm cat}<T_{S\to P}$. The later implies the following necessary conditions: $T_{C\:S\to C\.P_2\bs C\.S}<T_{S\to P}$, $T_{C\.P_2\to C\.P+P\bs C\:S}<T_{S\to P}$ and $T_{C\.P+P\to C+2P}<T_{S\to P}$. Note, however, that $T_{C+S \to C\.S}<T_{S\to P}$ is {\it not} a necessary condition, nor is  $T_{C\.S \to C\:S\bs C+S}<T_{S\to P}$, but while catalysis can occur for arbitrary values of $T_{C+S \to C\.S}$, the transition $C\.S \to C\:S$ is constrained as it enters $\rho_1$ which is itself part of $T_{\rm cat}$. This case applies to the problem of catalyzing the dissociation of a dimer outlined in the introduction~\cite{mrz}. It provides another example where a forward step along a catalytic path needs not be faster than the spontaneous reaction, here arising from the fact that the catalytic process is decomposed into a graph with several loops.

\subsection{Beyond single molecules}

We have considered so far an initial situation with a single catalyst and a single substrate. This condition is particularly interesting to design catalysts, as it has the lowest number of degrees of freedom, facilitating both calculations of first passage times and numerical simulations. This raises, however, question of whether this is necessarily the most favorable condition for catalysis. We provide an example showing that it is not necessarily the case: a molecule $C$ can accelerate the formation of the product in the presence of multiple substrates when it does not in the presence of a single one. We will see that for unimolecular reactions, such conclusion can be directly drawn from studying a single catalyst and a single substrate provided we go beyond mean first-passage times $T_{\rm cat}=\E[t_{\rm cat}]$ to consider the full distribution of first-passage times $\P[t_{\rm cat}]$. Moreover, the presence of additional molecules, either substrates, catalysts, or products, may also turn a catalyst into an inhibitor. As before, in this section we assume temperature and volume to be fixed so that the reaction rate constants do not change with time, and increasing the number of molecules increases their concentration.

\subsubsection{Necessary conditions for catalysis with multiple $S$ and $C$} \label{Conditions for catalysis with multiple $S$ and $C$}

\begin{figure}[t]
\begin{center}
\includegraphics[width=.5\linewidth]{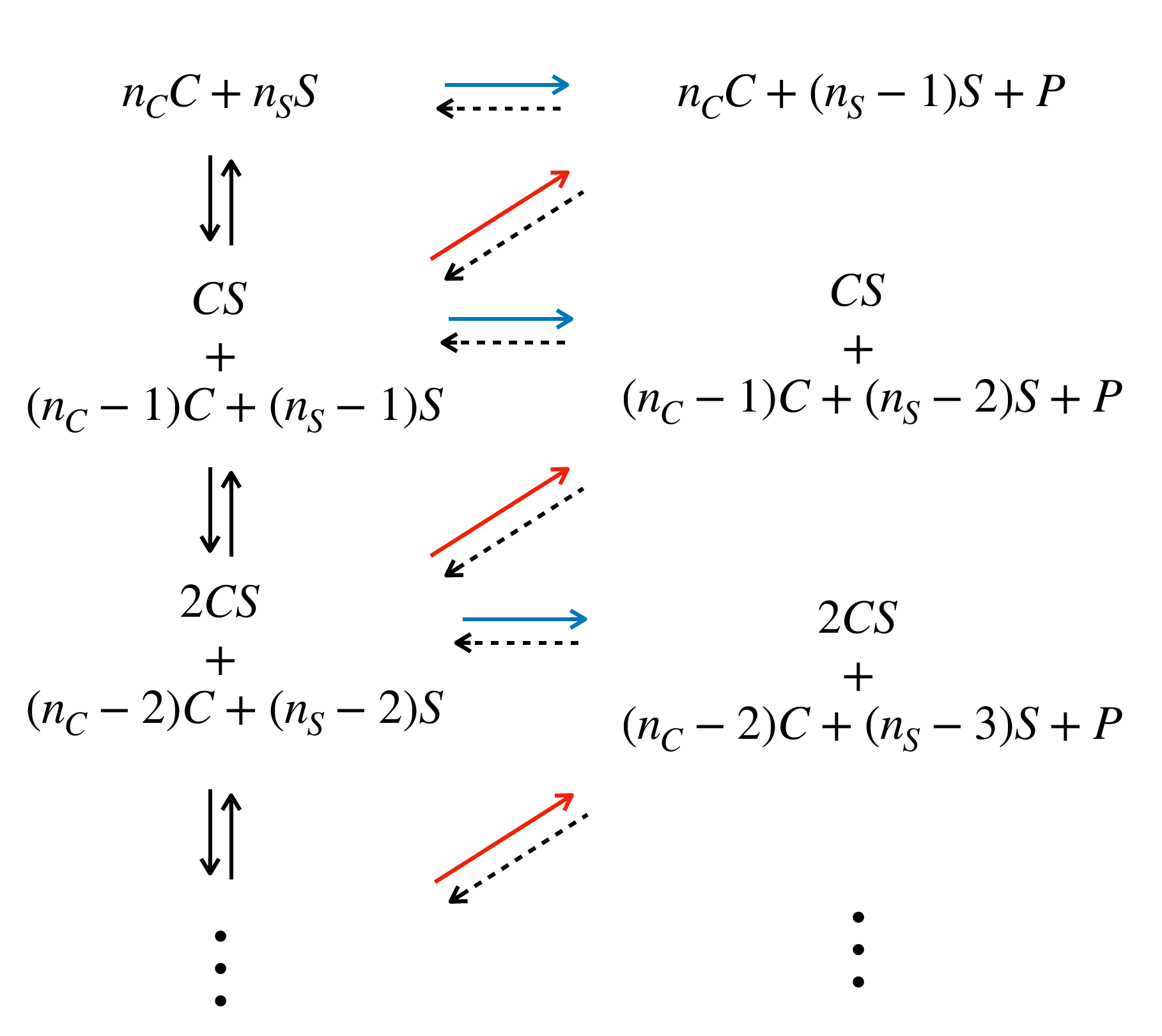}
\caption{Possible decomposition of the catalytic cycle for an unimolecular reaction starting with $n_s$ substrates $S$ and $n_c$ molecules $C$. In the extended definition of catalysis presented in the main text, the final state is any of the \confs on the right. Blue horizontal arrows represent spontaneous reactions, with the first one being the fastest of them all (with $n_s$ substrates). Red diagonal arrows represent catalytic reactions.} \label{fig: nsS_ncC}
\end{center}
\end{figure}

We first consider the substrate to be unimolecular and the reaction to proceed in a single step. For an initial condition consisting of $n_s$ substrates $S$, $n_c$ molecules $C$, and no product $P$, a generic decomposition of the catalytic cycle into elementary processes is represented in Fig.~\ref{fig: nsS_ncC}. As there are multiple substrates and catalysts in the system, one substrate can be transformed into a product while other substrates are either free, or interacting with catalysts. Thus, it is of interest to relax our original definition of catalysis and consider the mean time to form the first free product irrespectively of the other molecules, i.e., $T_{n_cC+n_sS\to X+P}$ where $X$ represents all molecules other than a product $P$. These are all of the states in the right column of Fig.~\ref{fig: nsS_ncC}. 
As this encompasses the case where all the molecules $C$ are \textit{not} interacting with other molecules -- as per our original definition -- the necessary conditions that we shall obtain under this extended definition also apply to the original one.

The possible final states on the right side of Fig.~\ref{fig: nsS_ncC} are reached through either a spontaneous reaction (horizontal blue arrows) or a catalyzed reaction (diagonal red arrows). Assuming the spontaneous reactions to be independent of each other, the mean time at which the first of $n_s$ spontaneous reactions is completed is $T_{n_sS\to P+(n_s-1)S}=\E[\min(t^{(1)}_{S\to P},\dots,t^{(n_s)}_{S\to P})]$ where $t_{S\to P}^{(q)}$ denotes a random variable for the time of completion of the spontaneous reaction in the presence of a single substrate, such that $T_{S\to P}=\E[t_{S\to P}^{(q)}]$ for any $q$. A necessary condition for catalysis is that one of the transitions from the left column to the right column in Fig.~\ref{fig: nsS_ncC} is, on average, faster than the spontaneous reaction with $n_s$ substrates (top blue arrow). Formally, this condition implies that there must exist an $r\geq 1$ for which
\beq\label{eq:catmult}
\E[\min(t^{(1)}_{S\to P},\dots,t^{(n_s-r)}_{S\to P},t^{(1)}_{\rm cat},\dots,t^{(r)}_{\rm cat})] < \E[\min(t^{(1)}_{S\to P},\dots,t^{(n_s)}_{S\to P})]
\eeq
where $t^{(q)}_{\rm cat}$ is the random variable whose mean is $T_{\rm cat}=\E[t^{(q)}_{\rm cat}]$.
This necessary criterion involves only the distributions of $t_{S\to P}$ and $t_{\rm cat}$ defined for a single substrate and a single catalyst. But as transitions within the first column of Fig.~\ref{fig: nsS_ncC} must also be considered, it is not a sufficient condition for catalysis.

\begin{figure}[t]
\begin{center}
\includegraphics[width=0.95\linewidth]{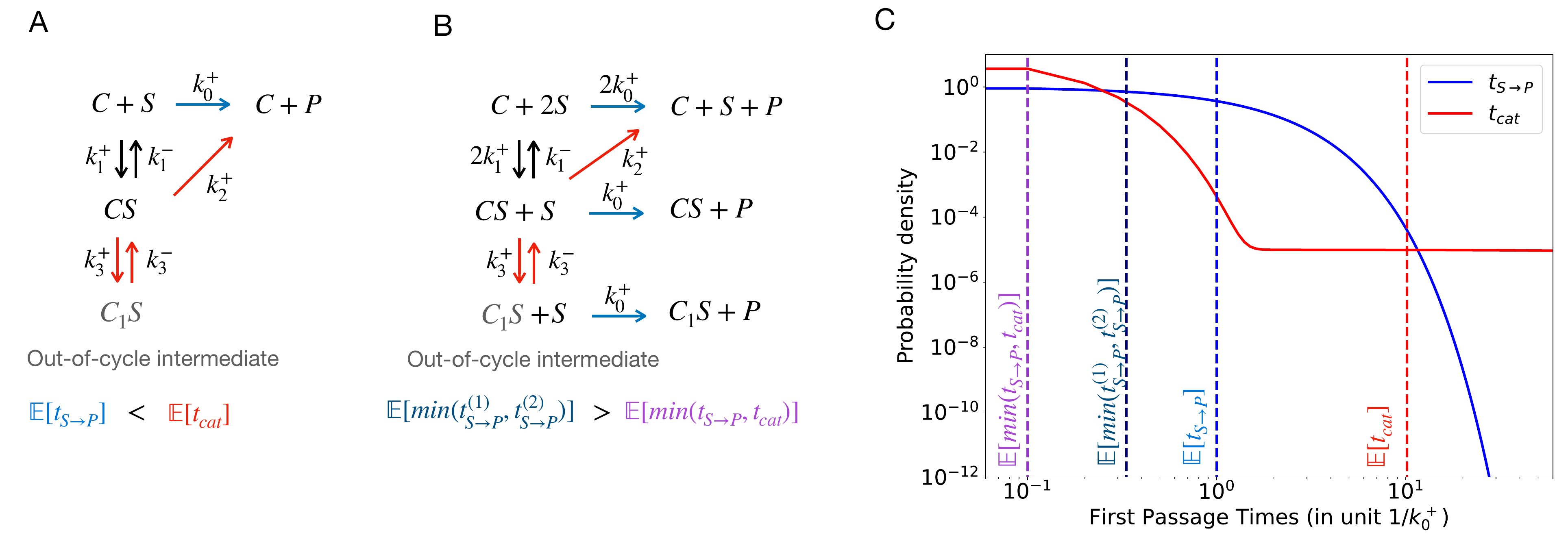}
\caption{Example of a molecule $C$ that catalyzes the formation of a product $P$ starting with two substrates $S$ but not with a single substrate. 
{\bf A.} We consider a situation where $k_3^-\ll k_3^+\ll k_2^+$ so that once a molecule $C$ has bound a substrate $S$, it has a small chance to trap the substrate in the long-lived \ys{out-of-cycle intermediate} $C_1S$. The mean time to form a product when the initial \conf contains only one substrate is longer in the presence of $C$ than that without it, $\E[t_{S\to P}]<\E[t_{\rm cat}]$, where $t_{S\to P}$ is the random variable for the time to complete the spontaneous reaction without catalyst and $t_{\rm cat}$ for reaching $C+P$ from $CS$ when a back transition to $C+S$ is excluded. {\bf B.} On the contrary, it is faster to form a product in the presence of $C$ if the system is initialized with two substrates, $\E[\min(t_{\rm cat}, t_{S\to P})]<\E[\min(t_{S\to P}^{(1)},t_{S\to P}^{(2)})]$. {\bf C.} In this case, the probability density $\P[t_{\rm cat}]$ is the sum of two exponentials, one controlled by $k_2^+$ that dominates at small times and the other controlled by $k_3^-$ that dominates at long times (Supporting Information \ref{Laplace}). As a consequence, the necessary condition for catalysis is fulfilled with two substrate molecules but not with one. The graph illustrates the case where $k_2^+/k_0^+=10$, $k_3^+/k_0^+=10^{-1}$ and $k_3^-/k_0^+=10^{-3}$.
}
\label{fig: not_exp}
\end{center} 
\end{figure}

The distributions $t_{S\to P}$ and $t_{\rm cat}$ are not necessarily exponential; if they are, however, the necessary condition Eq.~\eqref{eq:catmult} reduces to $((n_s-r)T_{S\to P}^{-1}+rT_{\rm cat}^{-1})^{-1}<(n_sT_{S\to P}^{-1})^{-1}$, i.e., $T_{\rm cat}<T_{S\to P}$, the necessary condition for catalysis in the presence of a single substrate. On the other hand, if the distributions are not exponential, a molecule $C$ may not be a catalyst in a single copy with a single substrate but becomes one when in $n_c$ copies in the presence of $n_s$ substrates. An illustration is provided by the Markovian catalytic cycle represented in Fig.~\ref{fig: not_exp}. In this example, we have $\E[t_{\rm cat}]>\E[t_{S\to P}]$ due to the presence of an inactive \ys{out-of-cycle intermediate} of the cycle that can trap the substrate $S$ with a small probability but for a long time. However, when the system is initialized with two substrates, we find $\E[\min(t_{\rm cat}, t_{S\to P})]<\E[\min(t_{S\to P}^{(1)},t_{S\to P}^{(2)})]$. This is because either the reaction proceeds very quickly through the catalyst, or, if it does not, with one of the two substrates binding to the catalyst and getting trapped in the \ys{out-of-cycle intermediate}, because the other substrate can still spontaneously react. This is therefore an example where the reaction is on average faster with $C$ in the reaction vessel when there are two substrates compared to when a single substrate is present.

\subsubsection{Efficiency of catalysis with multiple $S$, multiple $C$, or multiple $P$} \label{section: efficiency}

How the presence of multiple substrates, catalysts, or products affects catalysis also depends on the nature of the spontaneous reaction, as we illustrate here with three types of spontaneous reactions: with unimolecular substrate and product, with a bimolecular product, and with a bimolecular substrate. In order to develop intuition, we vary successively the number of substrates, the number of catalysts, or the number of products individually, while keeping the other two fixed. In this case, it is also easy to analytically compute the mean time to form the first product (see an example of formal derivation in Supporting Information \ref{app: multiple substrates}). We discuss the transformation of multiple substrates by multiple catalysts in the next section.

\begin{figure}[t]
\begin{center}
\includegraphics[width=1\linewidth]{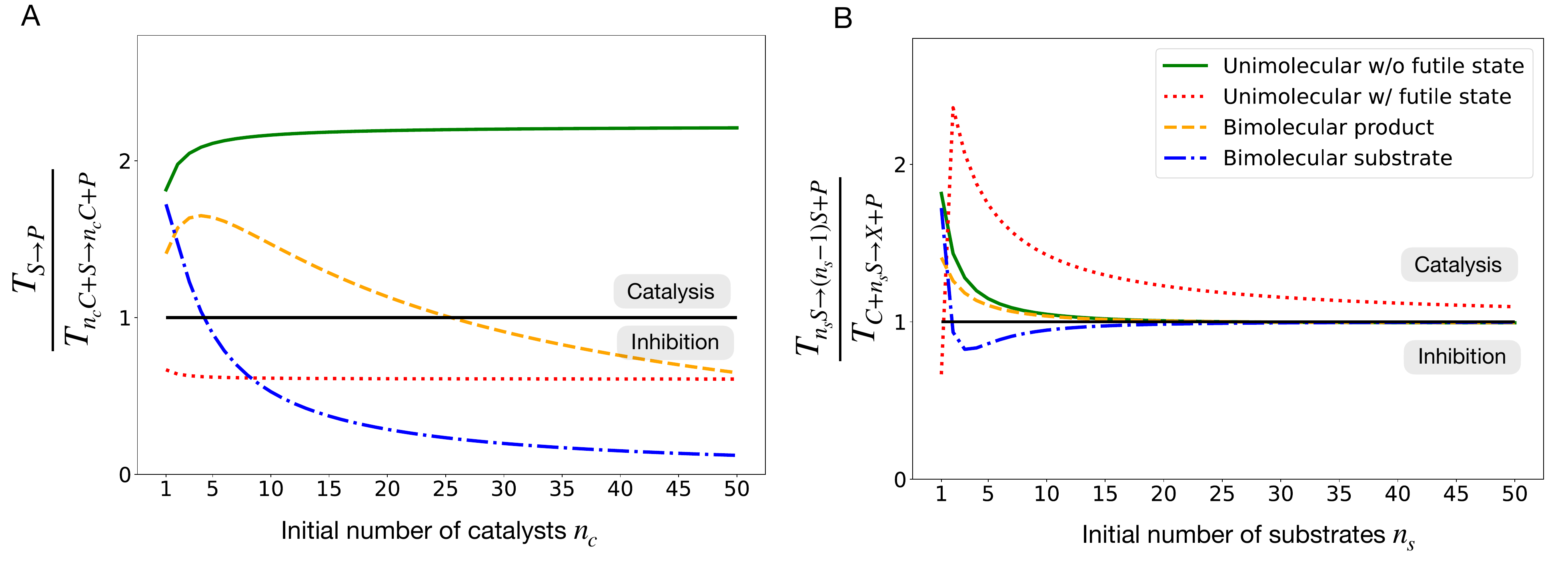}
\caption{Illustration of the effect of having multiple substrates or multiple catalysts in the reaction vessel on the efficiency of catalysis. We consider four types of spontaneous reactions: with a unimolecular substrate and product (green and red with no \ys{out-of-cycle intermediate} or with one, respectively), with a bimolecular product (yellow), and with a bimolecular substrate (blue). {\bf A.} Catalytic efficiency with one substrate ($n_s = 1$) and an increasing number of catalysts $n_c$. 
With unimolecular substrates, if $C$ is initially a catalyst, catalytic efficiency increases with the number of catalysts. For reactions with bimolecular products and/or substrates, adding more catalysts can be detrimental and eventually transforms a catalyst into an inhibitor ($T_{S\to P}/T_{n_cC+S\to n_cC+P}<1$). {\bf B.}  Catalytic efficiency with one catalyst ($n_c=1$) and an increasing number of substrates $n_s$. Here, we extend our original definition and compute the mean time to make the first product irrespective of whether catalysts are still interacting with other substrates , $T_{n_cC+n_sS\to X+P}$, where $X$ represents all molecules other than a product $P$. For all four types of reactions, in the limit of a large number of substrates, the products most likely arise through a spontaneous reaction. The values of the reaction rates are detailed in Supporting Information~\ref{app: multiple catalysts}. }
\label{fig: 4_table}
\end{center} 
\end{figure}

We start by considering the effect of increasing the number $n_c$ of catalysts in the presence of a single substrate, $n_s=1$. Examples of mean time to form the first product are shown in Fig.~\ref{fig: 4_table}A (see details in  Supporting Information \ref{app: multiple substrates}). For unimolecular reactions, adding more catalysts in the reaction vessel increases the probability that the substrate meets a catalyst before spontaneously reacting. Thus, if $C$ is a catalyst when $n_c=1$, then the catalytic efficiency increases with $n_c$, while if it is an inhibitor, then the efficiency decreases (Fig.~\ref{fig: 4_table}A). This is unlike the case of reactions with multi-molecular products (e.g., $S \to P + P$) or/and multi-molecular substrates (e.g., $S+S \to P$) where catalytic efficiency eventually decreases for large enough $n_c$, up to a point where any catalyst is turned into an inhibitor. When the product is multi-molecular, an excess of catalysts indeed causes any released molecule of the product ($P$) to bind to a catalyst, which hinders the joint release of all product's molecules ($2P$). With multi-molecular substrates, an excess of catalysts causes the different substrates ($2S$) to bind to different catalysts, which hinders their joint interaction with the same catalyst. As illustrated in Fig.~\ref{fig: 4_table}A, the dependence of catalytic activity on the number $n_c$ of catalysts can be non-monotonic as a small increase in the number of catalysts may favor the association of the substrate to a catalyst.

The inhibitory effects of large concentrations of catalysts are mitigated when the binding of different substrates and products to the catalyst is sequential (Supporting Information \ref{app: multiple catalysts}). Notably, when considering a single substrate ($n_s=1$), for none of the four reactions studied in Fig.~\ref{fig: 4_table} does a molecule $C$ that is not a catalyst in single copy ($n_c=1$) becomes one when present in multiple copies ($n_c>1$, see Supporting Information \ref{app: multiple catalysts}). This justifies studying a single catalyst in these cases.

Varying the number $n_s$ of substrates in the presence of a single catalyst ($n_c=1$) can also lead to non-monotonic effects. In the limit of a large number of substrates, the products most likely arise through a spontaneous reaction and the catalytic efficiency is marginal (Fig.~\ref{fig: 4_table}B).
Catalytic efficiency may vary non-monotonically with the number of substrate for different reasons; for example, because of a non-exponential distribution of $t_{\rm cat}$ (as in Fig.~\ref{fig: not_exp}), or because the spontaneous reaction scales non-linearly with the number of substrates (as for reactions with bimolecular substrates). We note that for none of the reactions without an \ys{out-of-cycle intermediate} studied in Fig.~\ref{fig: 4_table} does a molecule $C$ that is not a catalyst with a single substrate ($n_s=1$) become one in the presence of multiple substrates (for $n_s>1$, see Supporting Information \ref{app: multiple substrates}). This justifies studying a single substrate in these cases.

In the presence of multiple products, extra states must be considered in which molecules $C$ are bound to product molecules. Consistent with the general intuition that such product inhibition can only be detrimental, catalysis occurs in the presence of multiple products only if it occurs with no products (Supporting Information \ref{app: multiple P logic}). An initial absence of products is therefore always most favorable for catalysis and, as we noted earlier, it is also advantageous to remove products as they are produced.

\subsubsection{Catalysis of multiple substrates} \label{Section D}

\begin{figure}[t]
\begin{center}
\includegraphics[width=0.9\linewidth]{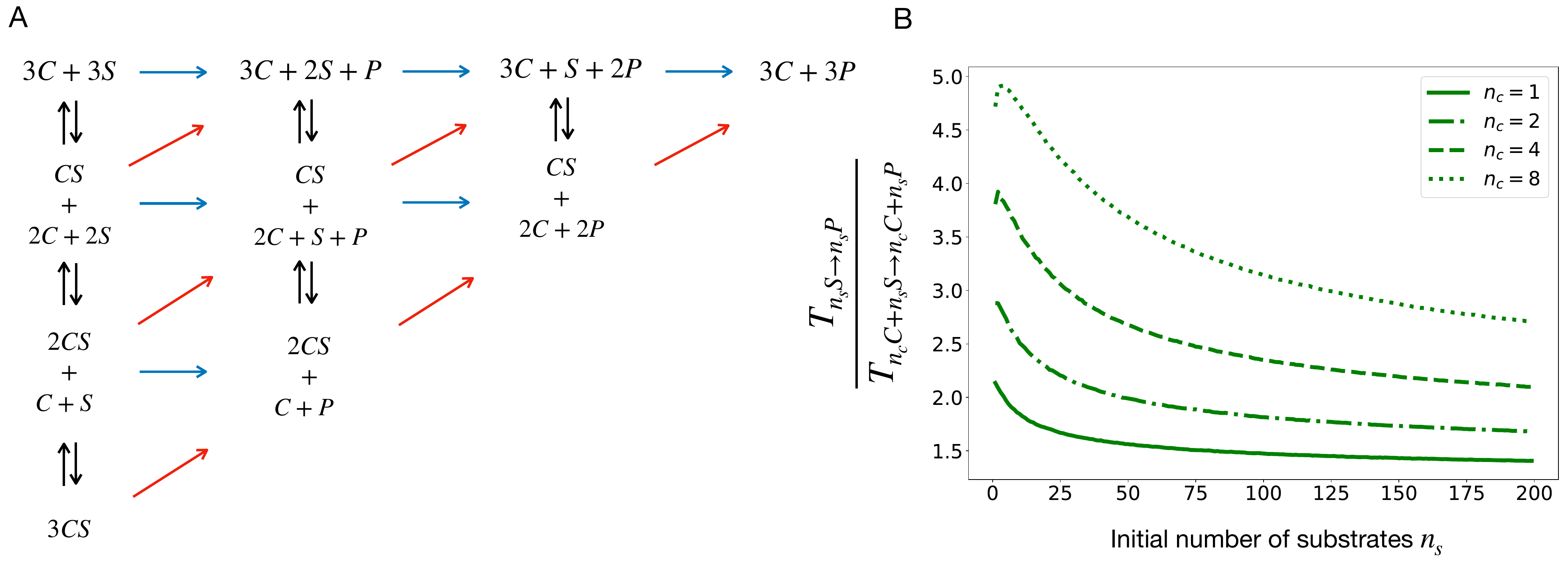}
\caption{{\bf A.} Extension of Fig.~\ref{fig: nsS_ncC} beyond the first product, represented here with $n_c = n_s = 3$ for clarity. The spontaneous reaction is assumed to be irreversible, and products are removed upon forming. {\bf B.} Gillespie simulations: mean time to form $n_s$ products from $n_s$ substrates and $n_c$ catalysts. We recover the same dependency on the number of substrates and catalysts as in Fig.~\ref{fig: 4_table} for the case of unimolecular reactions: as more substrate molecules need to be transformed, the catalytic efficiency eventually decreases, and adding more catalysts is always beneficial.} \label{fig: scheme 3 products}
\end{center} 
\end{figure}

We have focused so far on the time to make the first product, but the same principles apply when considering the time to make the first $n_p > 1$ products out of $n_s$ substrates that are initially present. Indeed, representing the catalytic cycle as in Fig.~\ref{fig: scheme 3 products}A, we can derive a series of necessary conditions of the form of Eq.~\eqref{eq:catmult} (see Supporting Information \ref{app: beyond first product}).
We find also the same patterns of dependency on the number of substrates and catalysts: 
as more substrate molecules need to be transformed, the catalytic efficiency eventually decreases, and, for unimolecular reactions, adding more catalysts is always beneficial, as illustrated in Fig.~\ref{fig: scheme 3 products}B.

\subsection{Relation to Michaelis-Menten kinetics}

Mean first-passage times are known to provide an alternative derivation of Michaelis-Menten kinetics in enzymology~\cite{ninio1987alternative,qian2008cooperativity}, which is useful to analyze single-molecule experiments~\cite{min2005fluctuating} or to understand theoretically what controls enzyme~\cite{banerjee2017accuracy}. We explain here how the mean first-passage time $T_{\MM}$ defined in this context is related but different from the mean first-passage time $T_{C+S\to C+P}$ introduced in this work.

This is most simply explained in the case where catalysis is described by a Markov chain with a single intermediate state $CS$, as considered in Sec.~\ref{sec:Markov}. In enzymology, this scheme is usually studied under the assumptions that the spontaneous reaction is negligible, in which case it is written $C+S\reac{k_{-1}}{k_1}CS\xrightarrow[]{k_2} C+P$. The concentration $[S]$ of substrates is also assumed to much larger than the concentration $[C]$ of catalysts. The focus is then on the rate of product formation given by
\beq\label{eq:MM}
v_{\MM}=\frac{d[P]}{dt}=\frac{\kcat [S][C]}{K_M+[S]}
\eeq
with $\kcat=k_2$ and $K_M=(k_{-1}+k_2)/\bar k_1$. Here $\bar k_1$ is the second-order rate constant of the transition $C+S\to CS$. $v_{\MM}$ can also be obtained as a mean first-passage time $T_{\MM}$ from $C+S$ to $C+P$ as $v_{\MM}=[C]/T_{\MM}$ when viewing $C+S\reac{}{}CS\xrightarrow[]{} C+P$ as describing the conversion into product of one of the many available substrates by one particular enzyme~\cite{ninio1987alternative}. In this context, $k_1$ is a pseudo-first-order rate constant giving the rate per catalyst at which the complex $CS$ is produced from $C+S$, related to $\bar k_1$ by $k_1=\bar k_1[S]$. 

In our approach, we take the point of the view of a substrate rather than the point of view of a catalyst, which allows us to make a comparison with the spontaneous reaction. In this context, $k_1=\bar k_1/V$ if considering a single catalyst, or $k_1=\bar k_1[C]$ if considering many more catalysts than substrates. Here $\bar k_1$ is the same second-order rate as above and $V$ denotes the volume of the reaction vessel. 

Comparing the expressions for $\kcat$ and $K_M$ in enzymology and those of $T_{\rm cat}$ and $\rho_{\rm cat}$ in our approach, we verify that $\kcat=1/T_{\rm cat}$ and that the probability of encountering $\rho_{\rm cat}$ is closely related to $\kcat/K_M$. Eq.~\eqref{eq:rhoN1} indeed indicates that $\rho_{\rm cat}^{-1}-1=[(k_{-1}+k_2)/k_1](k_0/k_2)=(K_M/\kcat)(k_0/V)$. To make a more general connection between the two approaches, we can rewrite Eq.~\eqref{eq:central} as
\beq
\frac{T_{S\to P}}{T_{C+S\to C+P}}=1+\frac{T_{S\to P}/T_{\rm cat}-1}{1+(\rho_{\rm cat}^{-1}-1)T_{S\to P}/T_{\rm cat}}.
\eeq
Assuming the spontaneous reaction to be negligible, $T_{S\to P}=1/k_0\ll T_{\rm cat}=1/k_2$, and 
using the expression for $\rho_{\rm cat}$, we obtain
\beq
\frac{1}{T_{C+S\to C+P}}\simeq \frac{k_2}{1+(k_{-1}+k_2)/k_1}=\frac{\kcat }{1+K_MV}.
\eeq
when considering a single substrate in the presence of a single catalyst, or $1/T_{C+S\to C+P}=\kcat [C]/(K_M+[C])$ when considering a single substrate in the presence of many catalysts.

Since $T_{\rm cat}=1/k_{\rm cat}$, our criterion for catalysis $T_{\rm cat}<T_{S\to P}$ reads $k_{\rm cat}>k_0$ in the language of Michaelis-Menten kinetics. The ratio  $\kcat/k_0$ used by Wolfenden and collaborators to compare a catalyzed reaction to a spontaneous one~\cite{wolfenden2001depth} therefore corresponds to $T_{S\to P}/T_{C+S\to C+P}$ in the saturation limit where $\rho_{\rm cat}\to 1$ or $K_M\to 0$.

\section{Conclusion}

Motivated by the computational design of a minimal catalyst in the realm of colloids with programmable interactions~\cite{mrz}, we studied several questions related to the kinetics of catalysis. First, we addressed the question of assessing the presence of catalysis, for which we propose to compare the mean first-passage times from substrate(s) to product(s) in absence and in the presence of the candidate catalyst: a substance is a catalyst if the reaction is faster on average, formally $T_{C+S\to C+P}<T_{S\to P}$. Second, we addressed the question of scoring the performance of the catalyst, for which we propose to consider the ratio $T_{S\to P}/T_{C+S\to C+P}$ which, in the presence of catalysis, must be larger than one, that is, $T_{S\to P}/T_{C+S\to C+P}>1$. These quantities depend not only on the intrinsic properties of the catalyst but also on the extrinsic conditions under which catalysis is analyzed. In particular, we illustrated in several examples how it depends on the presence of multiple substrates, products or/and catalysts. We showed that it is not sufficient to analyze a single substrate and a single catalyst, although notable exceptions exist, including unimolecular single-step reactions. We also identified conditions that are more favorable to catalysis: if a substance is not a catalyst under these conditions, it is not a catalyst under other conditions. They include considering an irreversible spontaneous reaction with an equivalent forward rate or removing every product as soon as it is formed.

We showed how necessary conditions for catalysis can be derived from the analysis of a decomposition of the catalytic cycle into elementary steps. In the simplest cases, this decomposition takes the form of a Markov chain. While overall the reaction must be completed faster in the presence of the catalyst, it is not always necessary for every forward step along the cycle to be faster than the spontaneous reaction. In particular, when considering a linear scheme for catalysis $C+S\reac{\r_{-1}}{\r_1}CS_1\reac{\r_{-2}}{\r_2}\dots \reac{\r_{-N}}{\r_N}CS_N\xrightarrow[]{\r_{N+1}}  C+P$ along with the spontaneous reaction $S\xrightarrow{k_0}P$, it is required that the forward rates $\r_i$ satisfy $\r_i>\r_0$ for $i\geq 2$ but no such constraint applies to $\r_1$. Constraints on forward rates can take more relaxed forms when the elementary steps are not organized into a single cycle but instead in a graph with multiple loops. This happens for instance when the spontaneous reaction involves multiple substrates or multiple products. Informative necessary conditions on catalysis can also be derived when the decomposition of the catalytic cycle is not described by a Markov chain, although not without restrictive assumptions.

While motivated by the design of catalysts for experiments in soft-matter physics, our approach involves only the kinetics of catalysis and therefore has a broader scope. As we have shown, it is closely related to, although different, from quantifications of catalysis used in enzymology~\cite{ninio1987alternative,qian2008cooperativity}. The main difference is that we focus on the fate of a substrate rather than on the fate of a catalyst. This point of view is required to account for the spontaneous reaction, even in the presence of a catalyst. While the spontaneous reaction is typically negligible in the context of enzymes, a reference to the spontaneous reaction is essential both to define catalysis and to study it in conditions where catalytic efficiency is either poor or unknown, as it is for instance the case in experiments to design non-enzymatic autocatalysts~\cite{Kiedrowski}. Our approach should therefore find applications to the design and study of catalysts beyond our original case study~\cite{mrz}.

\acknowledgments{We acknowledge funding from ANR-22-CE06-0037. This work has received funding from the European Unions Horizon 2020 research and innovation program under the Marie Sklodowska-Curie grant agreement No. 754387.}

\newpage

\begin{center}
    \textbf{REFERENCES}\\
\end{center}


\newpage

\section{Supporting Information}

\subsection{Markovian catalytic cycles} \label{app:Markov}


{The simplest catalytic cycle is Markovian, and comprises only one intermediate state denoted as $CS$, which is accessible from either the unbound states $C+S$ or $C+P$ and can also transition back to those states. Graphically, the cycle is represented as
\vspace{0.2cm}
\begin{equation}
\begin{tikzcd}[column sep=small]
C+S \arrow[dr, shift left, "k_1"]\arrow[rr, "k_0"] & & C+P\arrow[ll, shift left, "k_{-0}"]\arrow[dl, "k_{-2}"] \\
    & CS \arrow[ul, "k_{-1}"]\arrow[ur, shift left, "k_2"]
\end{tikzcd}
\end{equation}
\vspace{0.2cm}

\noindent with $\r_{+n}$ denoting the forward rate of an elementary reaction and $\r_{-n}$ its reverse rate. Given that we consider the mean first-passage time $T_{C+S\to C+P}$ from the initial state $C+S$ to the absorbing state $C+P$, the transitions with rates $k_{-0}$ and $k_{-2}$ can be ignored and an equivalent representation is 
\beq
C+P\xleftarrow{k_0}C+S\reac{\r_{-1}}{\r_1}CS\xrightarrow[]{\r_{2}}  C+P,
\eeq
where $C+P$ is repeated on both sides.\\

When in $C+S$, the mean time before any transition is $T_{C+S}=1/(k_0+k_1)$, the probability to transition to $CS$ is $p_{C+S\to CS}=k_1/(k_0+k_1)$ and the probability to transition directly to $C+P$ is $p_{C+S\to C+P}=1-p_{C+S\to CS}=k_0/(k_0+k_1)$~\cite{Statistical2010}. Similarly, when in $CS$, the mean time before any transition is $T_{CS}=1/(k_2+k_{-1})$, the probability to transition to $C+P$ is $p_{CS\to C+P}=k_2/(k_2+k_{-1})$ and the probability to transition back to $C+S$ is $p_{CS\to C+S}=k_{-1}/(k_2+k_{-1})$. In terms of these quantities, the mean first-passage time from $C+S$ to $C+P$, denoted $T_{C+S\to C+P}$, can be expressed as a function of the mean first-passage time from $CS$ to $C+P$, denoted $T_{CS\to CP}$ as~\cite{ninio1987alternative_sup, park_describing_2022}:
\bea
T_{C+S\to C+P}&=&T_{C+S}+p_{C+S\to CS}T_{CS\to C+P}\\
T_{CS\to C+P}&=&T_{CS}+p_{CS\to C+S}T_{C+S\to C+P},
\eea
i.e.,}
\bea
T_{C+S\to C+P}&=&\frac{1}{k_0+k_1}+\frac{k_1}{k_0+k_1}T_{CS\to C+P}\\
T_{CS\to C+P}&=&\frac{1}{k_1+k_2}+\frac{k_{-1}}{k_1+k_2}T_{C+S\to C+P},
\eea

Solving these equations lead to an explicit expression of $T_{C+S\to C+P}$ as a function of the elementary rates:
\beq\label{eq:support_two}
T_{C+S\to C+P}=\frac{k_1+k_{-1}+k_2}{k_0k_{-1}+k_0k_2+k_1k_2}.
\eeq
To highlight constraints, we can rewrite Eq.~\eqref{eq:support_two} as follows:
\beq\label{eq:support_central}
T_{C+S\to C+P}=T_{S\to P}+\rcat (T_{\rm cat}-T_{S\to P}),
\eeq
with $T_{S\to P}=1/k_0$, $T_{\rm cat}=1/k_2$ and
\beq\label{eq:support_rhoN1}
\rcat=\frac{1}{1+(1+k_{-1}/k_2)k_0/k_1},
\eeq
where $T_{\rm cat}$ denotes the mean time to form the product once the substrate is bound to the autocatalyst, while $\rcat$ denotes the probability to reach $C+P$ from $C+S$ through the catalytic pathway ($CS\to C+P$). Indeed, let $p_1=k_1/(k_0+k_1)$ be the probability to transition from $C+S$ to $CS$ and $p_2=k_2/(k_{-1}+k_2)$ the probability to transition from $CS$ to $C+P$. The probability to reach $C+P$ from $C+S$ through the $CS\to C+P$ is:
\beq
\rho_{\rm cat}=p_1\left(\sum_{n=0}^\infty(1-p_2)^np_1^n\right)p_2=\frac{p_1p_2}{1-(1-p_2)p_1}=\frac{1}{1+(1+k_{-1}/k_2)k_0/k_1}.
\eeq
where $n$ is the number of times the back transition $CS\to C+S$ occurs.

\subsection{General criterion for catalysis} \label{app: non-Markovian} \label{app:semiMarkov}

We present here the derivation of Eq.~\eqref{eq:support_central} for the general case where the spontaneous reaction $S\to P$ occurs at a given rate. A single $S$ and a single $C$ are considered. We assume that a state $C\.\.S$ can be defined as a set of configurations where $C$ and $S$ are not interacting and where the probability and the nature of their future interactions are equivalent. This state defines a boundary between interacting and non-interacting systems. The initial configuration of the system denoted $(C+S)_0$, is assumed to be a non-interacting configuration and the final configuration, as well the final configuration $C+P$. From $(C+S)_0$, the system may either reach $C\.\.S$ by diffusion or reach $C+P$ through the spontaneous reaction $S\to P$. Once in $C\.\.S$, $C$ and $S$ may either move apart to $C\.\.S^+$ or move closer to $C\.\.S^-$, where $C\.\.S^+$ and $C\.\.S^-$ represent configurations infinitesimally close to $C\.\.S$. From $C\.\.S^+$, the system may either come back to $C\.\.S$ or reach the absorbing state $C+P$ through a spontaneous reaction. From $C\.\.S^-$, on the other hand, it may either come back to $C\.\.S$ or reach the absorbing state $C+P$ through catalysis. We will compare this situation to a situation where $C\.\.S^+$ is inaccessible, which may for instance be implemented in numerical simulations by confining the system in a box with reflecting boundaries defined by $C\.\.S$: the mean time to reach $C+P$ from $C\.\.S$ in this case is what we define as  $T_{\rm cat}$ and also denote as $T_{C\.\.S\to C+P\bs(C+S)}$, where $\bs (C+S)$ indicates that the configurations $C+S$ are excluded.

To derive  Eq.~\eqref{eq:support_central} we make the following hypotheses: $S\to P$ occurs at a rate $k_0$, $(C+S)_0\to C\.\.S$ takes a mean time $T'_{(C+S)_0\to C\.\.S}$ and $C\.\.S^+\to C\.\.S$ a mean time $T'_{C\.\.S^+\to C\.\.S}$, both in absence of spontaneous reaction. Eq.~\eqref{eq:support_central} then results from combining the following three equations which are derived below.

\begin{figure}[t]
\begin{center}
\includegraphics[width=.5\linewidth]{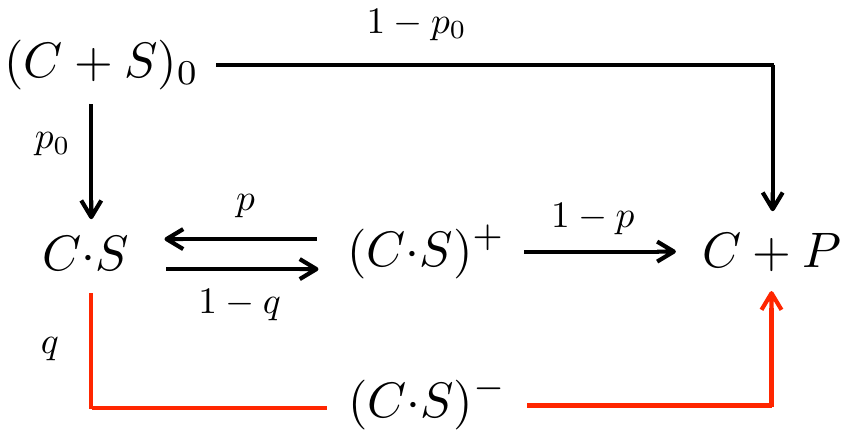}
\caption{States in the derivation of the necessary and sufficient condition on catalysis. We consider 3 intermediate states in addition to the initial state $(C+S)_0$ and the final absorbing state $C+P$. The probabilities to transition between these states are indicated next to the arrows. The times to transition from  $(C+S)_0$ to $C\.S$ given $p_0=1$, from $C\.S^+$ to $C\.S$ given $p=1$ and from $C\.S$ to $C+P$ given $q=1$ may follow arbitrary distributions, but the times to transition from $S$ to $P$ in absence of $C$ are assumed to be exponentially distributed. The time to transition from $C\.S$ to $(C\.S)^+$ given $q=0$ is assumed to be negligible. The mean time $T_{(C+S)_0\to C+P}$ to reach $C+P$ from $(C+S)_0$ is expressed in terms of the overall probability $\rho_{\rm cat}=p_0q/(1+(1-q)p)$ to take the catalytic (red) route and of the mean time $T_{\rm cat}$ to transition from $C\.S$ to $C+P$ given $q=1$ as $T_{(C+S)_0\to C+P}=T_{S\to P}+\rho_{\rm cat}(T_{\rm cat}-T_{S\to P})$ where $T_{S\to P}$ is the mean time to reach $P$ from $S$ in absence of $C$.\label{fig:ms}}
\end{center} 
\end{figure}

First, we have
\beq\label{eq:f1}
T_{(C+S)_0\to C+P}=p_0T_{C\.S\to C+P}+(1-p_0)T_{S\to P}
\eeq
where $p_0$ is the probability to reach $C\.S$ from $(C+S)_0$ before any spontaneous reaction  (Fig.~\ref{fig:ms}).

Second, we have 
\beq\label{eq:f2}
T_{C\.S\to C+P}=qT_{C\.S\to C+P\bs C+S}+(1-q)T_{C\.S^+\to C+P}
\eeq
where $q$ is the probability to reach $C+P$ from $C\.S$ without ever revisiting $C\.S^+$  (Fig.~\ref{fig:ms}). Here $T_{C\.S\to C+P\bs C+S}$ could also be written $T_{C\.S\to C+P\bs C\.S^+}$ and the assumption is that $T_{C\.S\to C\.S^+\bs (C\.S)^-}=0$.

Third, we have
\beq\label{eq:f3}
T_{C\.S^+\to C+P}=pT_{C\.S\to C+P}+(1-p)T_{S\to P}
\eeq
where $p$ is the probability to come back to $C\.S$ after an excursion in $C+S$  before any spontaneous reaction  (Fig.~\ref{fig:ms}).

Combining these three equations leads to
\beq
T_{(C+S)_0\to C+P}=\rho T_{C\.S\to C+P\bs C+S}+(1-\rho)T_{S\to P} \label{eq: total}
\eeq
which is equivalent to Eq.~\eqref{eq:support_central}, with
\beq\label{eq:rho}
\rho=\frac{p_0q}{1-(1-q)p}.
\eeq
To interpret $\rho$, note that it may also be written
\beq
\rho=\sum_{n=0}^\infty p_0(1-q)^n p^nq
\eeq
where the sum is over the probabilities that $(C+S)_0$ reaches $C\.S$ (factor $p_0$), ``unbind'' $n$ times to $C+S$ (factor $(1-q)^n$) and ``rebinds'' as many times to $C\.S$ (factor $p^n$) before following the catalytic route towards $C+P$ (factor $q$): $\rho$ therefore represents the overall probability to reach $C+P$ through catalysis when starting from $(C+S)_0$.

\subsubsection{Derivation of Eq.~\eqref{eq:f1}}

Consider that starting from $(C+S)_0$ the system takes a mean time $T_{(C+S)_0}$ before it either reaches $C\.S$, with probability $p_0$, or $C+P$, with probability $1-p_0$ so that
\beq
T_{(C+S)_0\to C+P}=T_{(C+S)_0}+p_0T_{C\.S\to C+P}.
\eeq
Eq.~\eqref{eq:f1} is then obtained by noting that $T_{(C+S)_0}=(1-p_0)T_{S\to P}$. To show this later relation, the essential ingredient is that the spontaneous reaction occurs within a time $t$ that is exponentially distributed with a rate $k_0$ such that $T_{S\to P}=1/k_0$. The time $\tau$ for $(C+S)_0$ to diffuse towards $C\.S$ in absence of any possible spontaneous reaction may, on the other hand, follow an arbitrary distribution $\chi(\tau)$. Under these assumptions, we have
\beq
p_0=\int_{0}^\infty d\tau\chi(\tau)\int_{\tau}^\infty dt ke^{-kt}=\int_{0}^\infty d\tau\chi(\tau) e^{-k\tau}=\langle e^{-k\tau}\rangle
\eeq
were $\langle\cdot\rangle$ denotes an average over $\tau$ based on $\chi(\tau)$, and
\bea
T_{(C+S)_0}&=&\int_0^\infty d\tau \chi(\tau) \int_0^\infty dt k_0e^{-k_0t}[t 1(t<\tau)+\tau 1(\tau<t)]\\
&=&\langle \int_0^\tau dt\ tk_0e^{-k_0t}+\tau \int_\tau^t dt\ k_0e^{-k_0t}\rangle=\frac{1}{k_0}\left(1-\langle e^{-k_0\tau}\rangle\right)
\eea
from which it follows that $T_{(C+S)_0}=(1-p_0)T_{S\to P}$.

Note that if $\tau$ is exponentially distributed with mean $1/k_1$ we have simply $p_0=k_1/(k_0+k_1)$ and $T_{(C+S)_0}=1/(k_0+k_1)$ which leads directly to $T_{(C+S)_0}=(1-p_0)T_{S\to P}$.

\subsubsection{Derivation of Eq.~\eqref{eq:f2}}

Let $\a$ be the probability that when in $C\.S$ the system goes towards $C\.S^-$ rather than towards $C\.S^+$ (in no time) and let $\beta$ be the probability that when in $C\.S^-$ the system will come back at least once to $C\.S$. By definition,
\bea
T_{C\.S\to C+P}&=&\a T_{C\.S^-\to C+P}+(1-\a)T_{C\.S^+\to C+P}\\
T_{C\.S^-\to C+P}&=&\beta(T_{C\.S^-\to C\.S}+T_{C\.S\to C+P})+(1-\beta)T_{C\.S^-\to C+P\bs C\.S}
\eea
where $T_{C\.S^-\to C+P\bs C\.S}$ denotes the time to reach $C+P$ from $C\.S^-$ without ever visiting $C\.S$. From these equations it follows that
\beq
T_{C\.S\to C+P}=\frac{1}{1-\a\beta}\left(\a(\beta T_{C\.S^-\to C\.S}+(1-\beta)T_{C\.S^-\to C+P\bs C\.S}))+(1-\a)T_{C\.S^+\to C+P}\right).
\eeq
The point is that $T_{\rm cat}=T_{C\.S\to C+P\bs C+S}$ can be written in the same way but with $\a=1$ since it correspond to a case where accessing $C\.S^+$ is excluded,
\beq
T_{\rm cat}=\frac{1}{1-\beta}\left[\beta T_{C\.S^-\to C\.S}+(1-\beta) T_{(C\.S^-\to C+P)\bs C\.S}\right]
\eeq
This leads to
\beq
T_{C\.S\to C+P}=\frac{1}{1-\a\beta}\left[\a(1-\b)T_{\rm cat}+(1-\a)T_{C\.S^+\to C+P}\right]
\eeq
which is equivalent to Eq.~\eqref{eq:f3} with 
\beq
q=\frac{\a(1-\beta)}{1-\a\beta}=\a\left(\sum_{n=0}^\infty\beta^n\a^n\right)(1-\beta)
\eeq
which can be interpreted as the probability to go from $C\.S$ to $C\.S^-$ (factor $\alpha$), come back $n$ times to $C\.S$ (factor $\beta^n$) and in each case immediately diffuse to $C\.S^-$ (factor $\a^n$) an arbitrary number $n$ of times (sum over $n$) before eventually reaching $C+P$ (factor $1-\beta$), i.e., $q$ is the probability to reach $C+P$ from $C\.S$ without ever visiting $C\.S^+$. 

\subsubsection{Derivation of Eq.~\eqref{eq:f3}}

The derivation of Eq.~\eqref{eq:f3} is essentially equivalent to the derivation of Eq.~\eqref{eq:f1} with a starting point $C\.S^+$ instead of $(C+S)_0$, with $p$ the probability to reach $C\.S$ replacing $p_0$ and with $\chi(\tau)$ now representing the distribution of times to reach $C\.S$ from $C\.S^+$ in absence of any possible spontaneous reaction. 

\subsection{Catalysis for dimer dissociation}\label{sec:dimer} \label{app: dimer dissociation}

We consider here the catalytic cycle represented in Fig.~\ref{fig:dimer}. To derive necessary and sufficient conditions, we start by applying Eq.~\eqref{eq:support_central} twice,
\bea
T_{C+S\to C+P}&=&(1-\rho_0)T_{S\to P}+\rho_0T_{C\.S\to C+P\bs C+S}\\
T_{C\.S\to C\.P\bs C+S}&=&(1-\rho_1)T_{S\to P}+\rho_1T_{C\:S\to C\.P\bs C\.S}
\eea
and then relate $T_{C\.S\to C+P\bs C+S}$ to $T_{C\.S\to C\.P\bs C+S}$ by introducing $\g=\P(C\.P\to C\.S\bs C+S)$:
\bea
T_{C\.S\to C+P\bs C+S}&=&T_{C\.S\to C\.P\bs C+S}+T_{C\.P\to C+P\bs C+S}\\
T_{C\.P\to C+P\bs C+S}&=&\g(T_{C\.P\to C\.S\bs C+S}+T_{C\.S\to C+P\bs C+S})+(1-\g) T_{C\.P\to C+P\bs C\.S}
\eea
i.e.,
\beq
T_{C\.S\to C+P\bs C+S}=\frac{1}{(1-\g)}\left(T_{C\.S\to C\.P\bs C+S}+\g T_{C\.P\to C\.S\bs C+S}+(1-\g)T_{C\.P\to C+P\bs C\.S}\right)
\eeq
All together,
\beq
T_{C+S\to C+P}=\left(1-\rho_0\frac{\rho_1-\g}{1-\g}\right)T_{S\to P}+\frac{\rho_0}{1-\g}\left(\rho_1T_{C\:S\to C\.P\bs C\.S}+\g T_{C\.P\to C\.S\bs C+S}+(1-\g)T_{C\.P\to C+P\bs C\.S}\right)
\eeq
which can also be written in the form $T_{C+S\to C+P}=(1-\rcat)T_{S\to P}+\rcat T_{\rm cat}$
with
\beq
\rcat=\rho_0\left(1-\frac{1-\rho_1}{1-\g}\right)=\rho_0\left(1-(1-\rho_1)\sum_{n=0}^\infty\g^n\right)
\eeq
that can be interpreted as the probability for the reaction not to occur spontaneously, neither through $C+S\to C+P$ nor through $C\.S\to C\.P$.

When $\g=0$, for instance because the reaction is irreversible, this simplifies to $\rcat=\rho_0\rho_1$ and $T_{\rm cat}=T_{C\:S\to C\.P\bs C\.S}+\rho_1^{-1}T_{C\.P\to C+P\bs C\.S}$ and a necessary condition for catalysis is therefore $T_{C\:S\to C\.P\bs C\.S}<T_{S\to P}$, which implies $T_{C\.S\to C\:P\bs C\.S}<T_{S\to P}$ and $T_{C\:P\to C\.P\bs C\:S}<T_{S\to P}$. On the other hand, there is no corresponding constraint on $T_{C\.S\to C\:S\bs C+S}$. Instead, a necessary condition on the transition $C\.S\to C\:S$ takes the form $\rho_1>T_{C\:P\to C\.P\bs C\:S}/T_{S\to P}$.

\subsection{Example of a catalytic cycle with an out-of-cycle intermediate} \label{Laplace}

Here we derive the distribution of the first passage times $\P[t_{\rm cat}]$ for the catalytic cycle with an out-of-cycle intermediate presented in Fig~\ref{fig: not_exp}, showing that when the distribution $t_{\rm cat}$ is not exponentially distributed, a molecule $C$ that is not a catalyst in a single copy with a single substrate may be a catalyst in a condition with $n_c \geq 1$ catalysts and $n_s \geq 1$ substrates. 

The catalytic scheme and notations for the states and rates are presented in Fig~\ref{fig: not_exp}A. $\P[t_{\rm cat}=t]=\partial P_{C+P}(t)/\partial t$ where $P_{C+P}(t)$ is obtained by considering the probabilities to be in each state at a function of time when starting at $t=0$ from $C_1S$ and when ignoring state $C+S$:
\begin{equation}
    \begin{cases}
    \partial_t P_{CS}(t) =- (k_2^+ + k_3^+)P_{CS}(t)  + k_3^-P_{C_1S}(t)\\
    \partial_t P_{C_1S}(t) =k_3^+ P_{CS}(t)  - k_3^-P_{C_1S}(t)\\
    \partial_t P_{C+P}(t) = k_2^+P_{CS}(t)\\
    \end{cases}
\end{equation}
with $P_{CS}(0) = 1$, $P_{C_1S}(0) = 0$, and $P_{C+P}(0)=0$.

The first two equations are decoupled from the third and are turned into a system of linear equation by considering the Laplace transforms $\hat P(s)=\int_0^\infty P(t)e^{-st}$ of the probability densities $P(t)$~\cite{ModernAnalysis1996}:
\begin{equation}
    \begin{cases}
    s\hat P_{CS}(s) -1 = - (k_2^+ + k_3^+)\hat P_{CS}(s)  + k_3^-\hat P_{C_1S}(s)\\
    s\hat P_{C_1S}(s) =k_3^+\hat P_{CS}(t)  - k_3^-\hat P_{C_1S}(t)
    \end{cases}
\end{equation}
Solving these equations and applying the inverse Laplace transform, we obtain $P_{C_1S}(t)$ and finally $\P[t_{\rm cat}=t]=k_2^+P_{CS}(t)$:

\begin{equation}
\P[t_{\rm cat}=t]=\frac{Ak_2^+}{2B}\left((-k_2^+ -k_3^+ + k_3^- + A)e^{-(k_2^+ +k_3^+ + k_3^- - A)t/2} - (-k_2^+ -k_3^+ + k_3^- - A)e^{-(k_2^+ +k_3^+ + k_3^- + A)t/2}\right)
\end{equation}
where $A = \sqrt{-4k_2^+k_3^- + (k_2^+ +k_3^+ + k_3^-)^2}$ and $B = (k_2^+)^2 + (k_3^+)^2 +  (k_3^-)^2 + 2k_2^+k_3^+  - 2k_2^+k_3^- + 2k_3^+k_3^-$. This is the probability density represented in Fig~\ref{fig: not_exp}B.

In the limit $k_3^-\ll k_2^+,k_3^+$, it simplifies to
\beq
\P[t_{\rm cat}=t]\simeq k_2^+\left(e^{-(k_2^++k_3^+)t}+\frac{k_3^-k_3^+}{(k_2^++k_3^+)^2}e^{-k_2^+k_3^-t/(k_2^++k_3^+)}\right)
\eeq
and if further assuming $k_3^+\ll k_2^+$,
\beq
\P[t_{\rm cat}=t]\simeq k_2^+\left(e^{-k_2^+t}+\frac{k_3^-k_3^+}{(k_2^+)^2}e^{-k_3^-t}\right)
\eeq
This shows that $\P[t_{\rm cat}]$ has two time scales: $1/k_2^+$ at short times and $1/k_3^-$ at long times.

\subsection{Conditions for catalysis in the presence of multiple catalysts or multiple substrates}\label{app: multiple substrates}\label{app: multiple catalysts}

\begin{figure}[t]
\begin{center}
\includegraphics[width=1\linewidth]{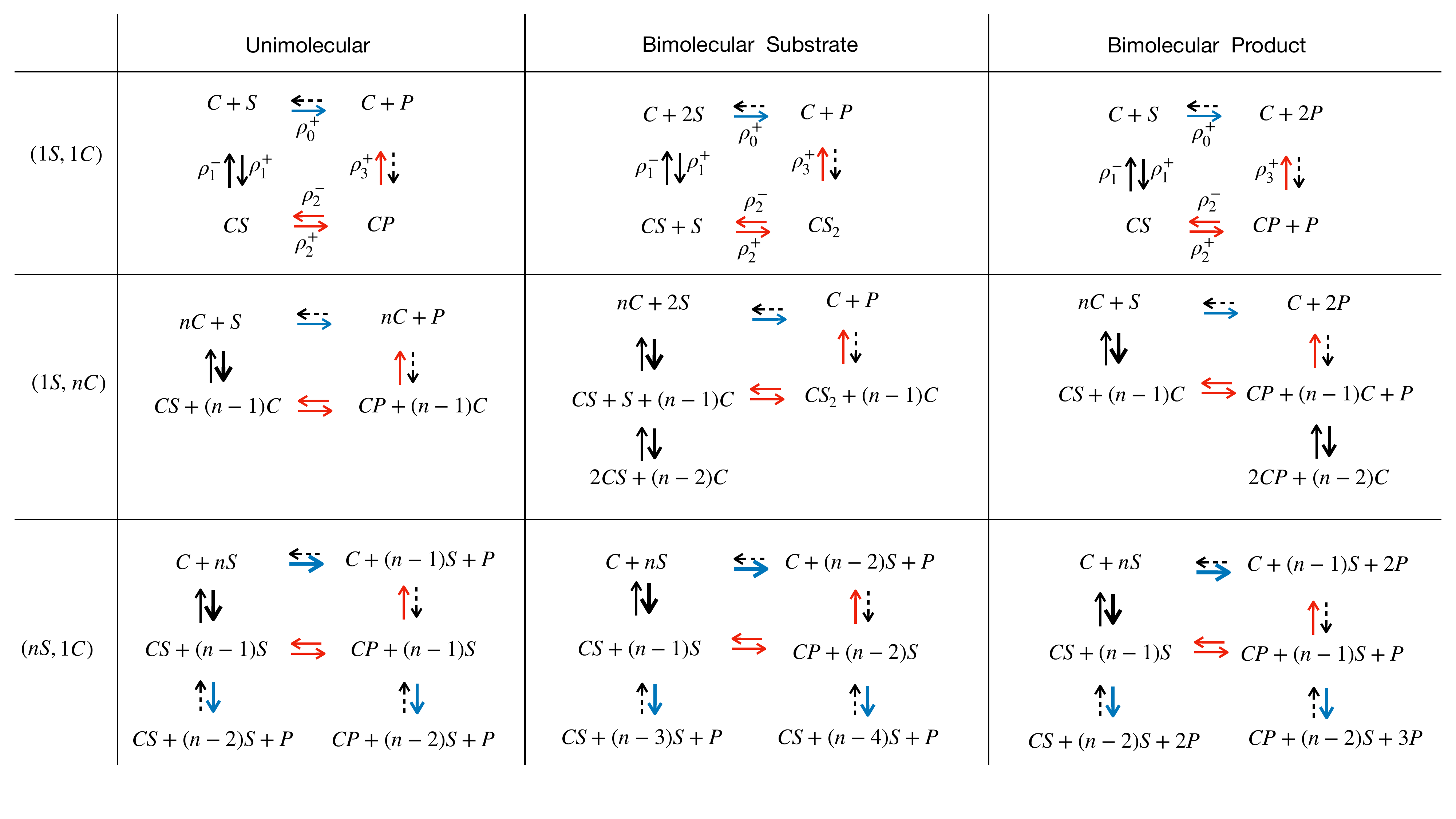}
\caption{Catalytic cycles for three particular reactions: a unimolecular reaction, a reaction with a bimolecular substrate, and another with bimolecular products. In the first row, we initiate the systems with one substrate and one catalyst; in the second row, with one substrate and $n$ catalysts; in the third row, with one catalyst and $n$ substrates. The rate $\rho_i$ for each transition $i$ is the product of the constant reaction rate $k_i$ and the numbers of reactant(s), e.g., $\rho_1^+=k_1^+n[C][S]$.\label{fig:nCnS} \label{fig: fig9}
}
\end{center} 
\end{figure}

We consider here the particular case where either the substrate or the catalyst is in a single copy and show that for three elementary reactions catalysis occurs in the presence of a single substrate and multiple catalysts or in the presence of a single catalyst and multiple substrates \textit{only if} it occurs with a single substrate and a single catalyst. The three reactions are a unimolecular reaction, a reaction with a bimolecular substrate, and another with bimolecular products. We assume a decomposition of the catalytic cycles as in Fig.~\ref{fig:nCnS}, and  further assume them to be Markovian. The rate $\rho_i$ for each transition $i$ is the product of the constant reaction rate $k_i$ and the numbers of reactant(s), e.g., $\rho_1^+=k_1^+n[C][S]$. Importantly, the conclusion obtained for these reactions does {\it not} extend to any spontaneous reaction, as the counter-example of Fig.~\ref{fig: not_exp} shows.

The conclusion can be drawn from the inspection of Fig.~\ref{fig:nCnS} when noting that the introduction of multiple catalysts (second row) or multiple substrates (third row) effectively modifies the catalytic cycle with a single substrate and a single catalyst (first row) in three possible ways. By modifying some of the rates (represented by the black arrows of large size), by adding additional states, or by effectively increasing the rate of the spontaneous reaction (when increasing the number of $S$). Crucially, however, the added states are only increasing the time to complete the catalytic route, and the modified rate in the main cycle are confined to $\rho_0^+$ to $\rho_1^+$ but $\rho_0^+$ only increases with the number of substrates, which effectively makes catalysis harder, while $\rho_1^+$ is involved in the efficiency of catalysis but not in the criterion for the presence of catalysis (see section \ref{section: General Criterion}). These observations imply that in all three cases catalysis occurs in the presence of a single substrate and multiple catalysts or in the presence of a single catalyst and multiple substrates only if it occurs with a single substrate and a single catalyst.

To demonstrate it more formally in one example, consider for instance the case of the reaction with a bimolecular substrate when considering a varying number $n$ of catalysts $C$ and a single substrate $S$ (central box in Fig.~\ref{fig:nCnS}). In this case, $\rho_1^+$ is effectively multiplied by a factor $n$ and an extra (futile) state has to be considered where two different catalysts are bound by one substrate each. The transition to this out-of-cycle intermediate $2CS + (n-2)C$ from $CS+S+(n-1)S$ occurs at rate $\rho_{01}^+ = \rho_1^+(n-1)$ and release from this state at a rate $\rho_{01}^- = 2\rho_1^-$. The condition for catalysis, $T_{n_cC + S \rightarrow n_cC + P} <T_{S \rightarrow P}$, can be derived by solving the system of equations associated with the reaction scheme (as in Section \ref{app:Markov}). It leads to
\begin{equation}
 \frac{1}{\rho_2^+} + \frac{1}{\rho_3^+} + \frac{\rho_2^-}{\rho_2^+\rho_3^+} + \frac{\rho_{01}^+(\rho_{2}^- + \rho_{3}^+)}{\rho_{01}^-(\rho_{2}^+\rho_{3}^+)}< \frac{1}{\rho_0^+} 
\end{equation}
We thus verify that $\rho_1^+$ is not involved and that the presence of an out-of-cycle intermediate makes the criterion for catalysis more stringent compared to the case with a single catalyst for which this criterion is simply
\begin{equation}
 \frac{1}{\rho_2^+} + \frac{1}{\rho_3^+} + \frac{\rho_2^-}{\rho_2^+\rho_3^+} < \frac{1}{\rho_0^+} 
\end{equation}
where the rates $\rho_2^+$, $\rho_2^-$ and $\rho_3^+$ are unchanged. Similar derivations can be done for the other cases.

We have considered here bimolecular reactions where the substrates or products are indistinguishable. If they are distinct and if their binding occur sequentially, i.e., $C+S_1+S_2 \leftrightarrow CS_1 \leftrightarrow CS_1S_2 \leftrightarrow C+P$ or $C+S \leftrightarrow CS \leftrightarrow CP_1 + P_2 \leftrightarrow C + P_1 + P_2$, then no out-of-cycle intermediates are present.

In Fig.~\ref{fig: 4_table}, we computed the mean-first passage time for each reaction scheme in Fig.~\ref{fig: fig9} and used the following reaction rates: for the unimolecular reaction with no out-of-cycle intermediate, $k_0^+ = 0.15$, $k_1^+ = 2$, $k_1^- = 1$, $k_2^+ = 1$, $k_2^- = 1$, and $k_3^+ = 1$;  for the reaction with a bimolecular product, $k_0^+ = 0.15$, $k_1^+ = 0.5$, $k_1^- = 1$, $k_2^+ = 1$, $k_2^- = 1$, $k_3^+ = 1$, and $k_3^+ = 0.3$; for the reaction with a bimolecular substrate, $k_0^+ = 0.08$, $k_1^+ = 0.1$, $k_1^- = 1$, $k_2^+ = 1$, $k_2^- = 1$, and $k_3^+ = 1$. For the unimolecular reaction with an out-of-cycle intermediate, we set $k_0^+ = 0.15$, $k_2^+ = 1$, $k_3^+ = 0.001$, $k_3^- = 0.0001$.

\subsection{Catalysis in the presence of multiple products} \label{app: multiple P logic}

\begin{figure}[t]
\begin{center}
\includegraphics[width=.3\linewidth]{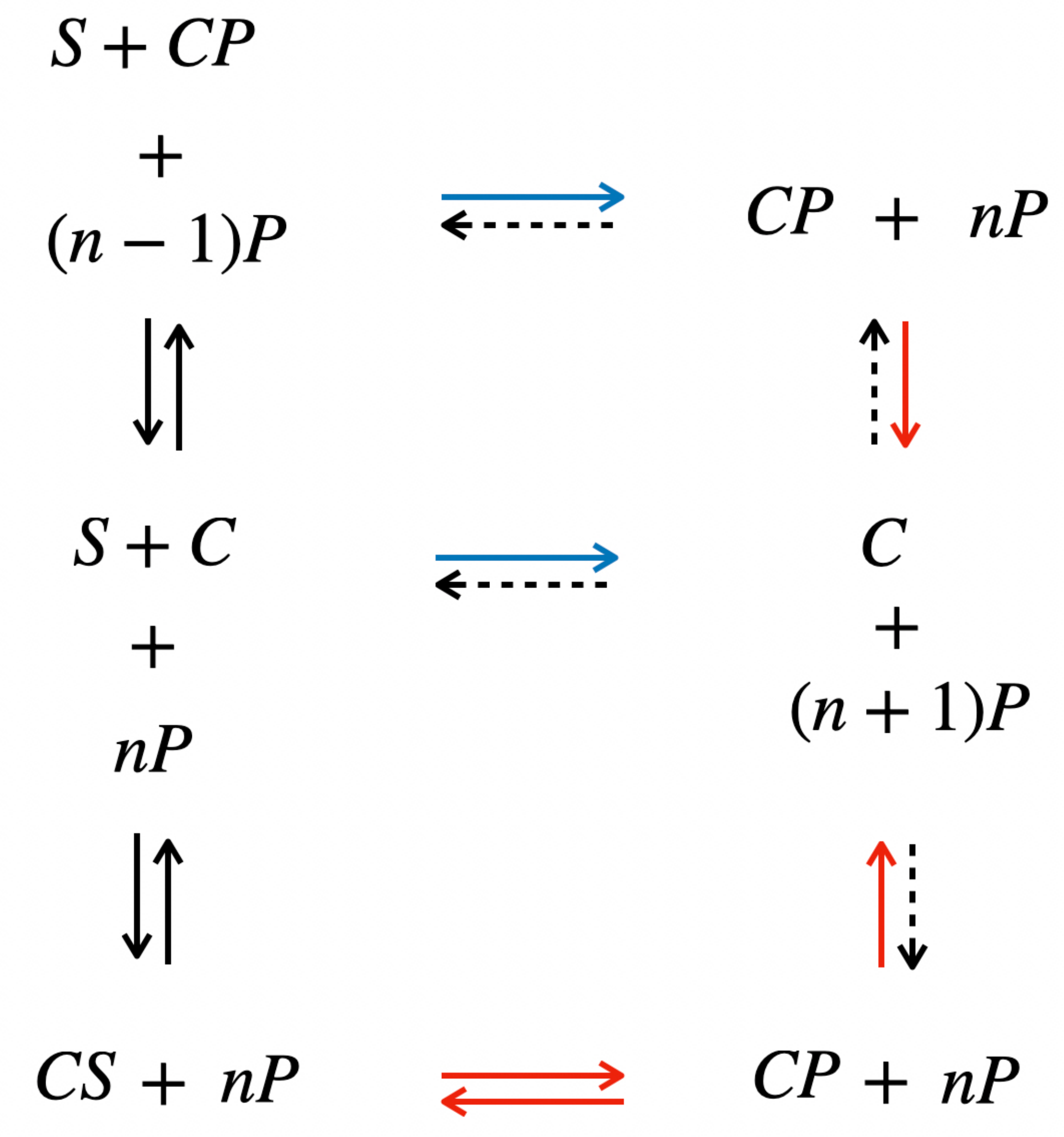} 
\caption{Catalytic cycle with multiple products for a unimolecular reaction $S\to P$ initiated with a single substrate, a single catalyst, and $n$ products. The difference with the case of no product ($n=0$) is the upper path that includes the spontaneous reaction together with the binding and unbinding of a product to the catalyst.\label{fig: n products}}
\end{center} 
\end{figure}

To illustrate how catalysis in the presence of multiple products requires catalysis to take place in the presence of no product, consider for instance the unimolecular reactions $S\to P$ represented in Fig.~\ref{fig: n products}, with initially one substrate $S$, one catalyst $C$ and $n$ products $P$. The final state of interest is $C + (n+1)P$. This state is reached either through the spontaneous reaction, the catalyzed reaction (path at the bottom) or through the upper path, which corresponds to the spontaneous reaction plus the binding/unbinding of the catalyst to a product. 

The conclusion follows from a simple observation: the presence of products only opens an addition path (the upper path of Fig.~\ref{fig: n products}), in addition to the spontaneous (middle) and catalytic (lower) paths that are present in absence of the product. As this additional path includes the spontaneous reaction together with the binding and unbinding of the catalyst to a product, it cannot be faster than the spontaneous reaction. As a consequence, catalysis must rely on the lower path, which is the same as in absence of products. $T_{\rm cat} < T_{S \rightarrow P}$ is therefore necessary for catalysis irrespectively of the number $n$ of products.

\subsection{Catalysis beyond the first product} \label{app: beyond first product}

We have represented in Fig.~\ref{fig: scheme 3 products}A the formation of $3$ products from $3$ substrate molecules with $3$ catalysts, when products are removed upon formation, and in the context of an irreversible spontaneous reaction. In this scheme, we see that the dynamics essentially boils down to moving from one column to the one on its right. We have shown that this can be faster than the corresponding spontaneous reactions (reactions on the top of the columns) only if there is at least one \conf on a left column that departs to a \conf on the right faster than the spontaneous reaction (Eq.~\eqref{eq:catmult}). A sufficient condition for catalysing $n_p$ reaction is that each of these moves is faster that their corresponding spontaneous reaction. A necessary condition is that the mean time for $n_p$ spontaneous reactions to proceed is longer than a combination of moves:\\
\begin{align*} \label{eq:catmult_multiple_products}
    &\E[\min(t^{(1)}_{S\to P},\dots,t^{(n_s-r)}_{S\to P},t^{(1)}_{\rm cat},\dots,t^{(r)}_{\rm cat})] + \dots + \E[\min(t^{(1)}_{S\to P},\dots,t^{(n_s-r-n_p)}_{S\to P},t^{(1)}_{\rm cat},\dots,t^{(r-n_p)}_{\rm cat})]\\
    &< \E[\min(t^{(1)}_{S\to P},\dots,t^{(n_s)}_{S\to P})] + \dots + \E[\min(t^{(1)}_{S\to P},\dots,t^{(n_s-n_p)}_{S\to P})]
\end{align*}
However, we note that even in the simplest case where products are systematically removed whenever produced, the total time $T_{n_sS+n_cC\to (n_s-n_p)S+n_cC+n_pP}$ {\it cannot} be computed as a sum $\sum_{r=0}^{n_p}T_{(n_s-r)S+n_cC\to (n_s-n_p)S+n_cC+P}$.

\end{document}